\documentclass[aps,prb,twocolumn,showpacs,10pt]{revtex4-1}

\usepackage{latexsym}
\usepackage{graphicx}
\usepackage{times,psfrag,subfigure}
\usepackage{amsmath}
\usepackage{dcolumn}
\usepackage{color}
\usepackage{latexsym,amsmath,amssymb,bm,euscript}
\usepackage{amssymb,amsmath,amsfonts,hyperref}
\usepackage{wasysym}
\usepackage{latexsym}
\usepackage{subfigure}

\begin{document}
\title{Spin-orbit coupling in $d^{2}$ ordered double perovskites}
\date{\today}

\author{Gang Chen}
\affiliation{Department of Physics, University of Colorado, Boulder, CO 80309}
\affiliation{JILA, University of Colorado, Boulder, CO 80309}
\author{Leon Balents}
\affiliation{Department of Physics,
University of California, Santa Barbara, CA 93106}
\affiliation{Kavli Institute for Theoretical Physics,
University of California, Santa Barbara, CA 93106}

\begin{abstract}

  We construct and analyze a microscopic model for insulating rock
  salt ordered double perovskites, with the chemical formula
  A$_2$BB'O$_6$, where the magnetic ion B' has a 4d$^2$ or 5d$^2$
  electronic configuration and forms a face centered cubic (fcc)
  lattice.  For these B' ions, the combination of the
  triply-degenerate antisymmetric two-electron orbital states and
  strong spin-orbit coupling forms local quintuplets with an effective
  spin moment $j=2$.  Moreover, due to strongly orbital-dependent
  exchange, the effective spins have substantial biquadratic and
  bicubic interactions (fourth and sixth order in the spins,
  respectively).  This leads, at the mean field level, to a rich
  ground state phase diagram which includes seven different phases: a
  uniform ferromagnetic phase with an ordering wavevector ${\bf p} =
  {\bf 0}$ and uniform magnetization along $[111]$ direction, four
  two-sublattice phases with an ordering wavevector ${\bf p} =
  2\pi(001)$ and two four-sublattice antiferromagnetic phases.
  Amongst the two-sublattice phases there is a quadrupolar ordered phase which
  preserves time reversal symmetry.  Extending the mean field theory
  to finite temperatures, we find ten different magnetization
  processes with different magnetic thermal transitions. In
  particular, we find that thermal fluctuations stabilize the
  two-sublattice quadrupolar ordered phase in a large portion of phase
  diagram.  Existing and possible future experiments are discussed in
  light of these theoretical predictions.

\end{abstract}
\date{\today}


%\email{gang.chen@colorado.edu}

\maketitle

%%%%%%%%%%%%%%%%%%%%%%%%%%%%%%%%%%%%%%%%%%%%%%%%%%%%%%%%%%%%%%%%%%%%%%%%%%%%%%%%%%%
\section{Introduction}
\label{sec:sec1}

The combination of strong electron correlation and strong
spin-orbit coupling (SOC) is relatively unexplored theoretically.  It
arises naturally in a broad family of magnetic Mott insulating systems
in which the three fold degenerate $t_{2g}$ orbitals are partially
filled. In these systems, the $t_{2g}$ orbital degeneracy is protected
by cubic lattice symmetry and the crystal field splitting is large
enough so that $e_g$ orbitals are not occupied.  Unlike for $e_g$
states, SOC is unquenched for $t_{2g}$ orbitals and splits the one
electron levels into an upper $j=1/2$ doublet and a lower $j=3/2$
quadruplet.\cite{PhysRevB.78.094403,kim2009} 

In this category, many
Ir-based magnets have been studied both theoretically and
experimentally.\cite{PhysRevB.78.094403,PhysRevLett.101.197202,pesin2010,PhysRevLett.99.137207,JPSJ.76.043706,JPSJ.71.2578,PhysRevLett.96.087204,kim2009}
Here, the magnetic ion Ir$^{4+}$ has a $d^5$ electron configuration
with five electrons residing on the $t_{2g}$ orbitals and the
effective $j=1/2$ description can be adopted. Some very exotic states
like a quantum spin liquid in
Na$_4$Ir$_3$O$_8$\cite{PhysRevB.78.094403,PhysRevLett.101.197202} and
a topological Mott insulator in
A$_2$Ir$_2$O$_7$\cite{pesin2010,JPSJ.76.043706,JPSJ.71.2578,PhysRevLett.96.087204}
have been observed and proposed in the strong coupling and
intermediate coupling regime, respectively.  

Moving beyond iridates, in our recent work,\cite{PhysRevB.82.174440}
we have studied the magnetic properties of a series of compounds
called ordered double perovskites with the chemical formula
A$_2$BB'O$_6$.\cite{stitzer:sss2002,wiebe:prb2002,wiebe:prb2003,yamamura:jssc2006,PhysRevLett.104.177202,PhysRevB.81.224409,PhysRevLett.99.016404}
We considered double perovskites in which the magnetic ions B' (e.g. Re$^{6+}$, Os$^{7+}$,
Mo$^{5+}$) have a $d^1$ electron configuration with one electron
residing on the $t_{2g}$ orbitals and hence form $j=3/2 $ local
moments.  In this analysis, we found several exotic phases including a
novel ferromagnetic state driven primarily by orbital interaction, an
antiferromagnetic state with strong octupolar order and a spin nematic
state, and furthermore a quantum spin liquid state postulated in a
region of the phase diagram.  

To round out the list of of magnetic systems with partially filled $t_{2g}$
orbitals, we must consider the $d^{2}$, $d^3$ and $d^4$ cases.  For a
$d^3$ electron configuration, the three electrons fill all the three
single-electron $t_{2g}$ orbitals, forming an antisymmetric orbital
wavefunction. The orbital degree of freedom is completely
quenched. The system is described by spin-only Hamiltonian with spin
$S=3/2$. Since it has a large spin, one may expect it to behave rather
classically.\cite{PhysRevB.80.134423} For a $d^4$ electron
configuration, when the SOC dominates over the Hund's coupling the
four electrons completely fill the lower $j=3/2$ quadruplets and there
is no local moment description at lowest order of approximation. When
the Hund's coupling dominates over SOC, the four $t_{2g}$ electrons
form a total spin $S=1$ but still have a three fold orbital
degeneracy. The effective SOC further lifts the spin-orbital
degeneracy completely and also favors a trivial $j=0$ local state.  So
the only non-trivial case left is the $d^2$ configuration, with two
electrons filling the $t_{2g}$ orbitals.  

In this paper, we consider this valence state in the context of double
perovskites mentioned above, specifically extending the theory of
Ref.~\onlinecite{PhysRevB.82.174440} to the case of a B' ion with a
$4d^2$ or $5d^2$ electron
configuration.\cite{PhysRevB.81.064436,yamamura:jssc2006} Unlike for
the $4d^1$ or $5d^1$ system with one electron per site, the local
Coulomb interaction plays an important role in determining the the
local spin and orbital structures for the $4d^2$ or $5d^2$ systems.
For the spin sector, the first Hund's rule requires a symmetrized spin
wavefunction, favoring a {\sl total spin} $S=1$. For the orbital
sector, the orbital electron wavefunction should be antisymmetrized,
composed of two single-electron $t_{2g}$ orbitals. These three
antisymmetrized two-electron states act as an effective $l=1$ {\sl
  total orbital angular moment}.  The strong SOC combines the total
spin $S=1$ with the total orbital moment $l=1$ and leads to an
effective total angular momentum $j=2$ description of the system.
Similar to the $j=3/2$ case studied in
Ref.~\onlinecite{PhysRevB.82.174440}, the orbitally-dependent nature
of the spin interactions lead to an interesting microscopic
Hamiltonian which contains significant biquadratic (fourth order in
spin operators) and triquadratic (sixth order in spin operators)
spin-spin interactions.  These unusual interactions may be understood
as couplings between the local magnetic quadrupole and octupole
moments. Therefore, an analysis of the microscopic Hamiltonian
naturally leads to a rich structure of magnetic multipolar orders.  

\begin{figure}[htp]
\includegraphics[width=8.0cm]{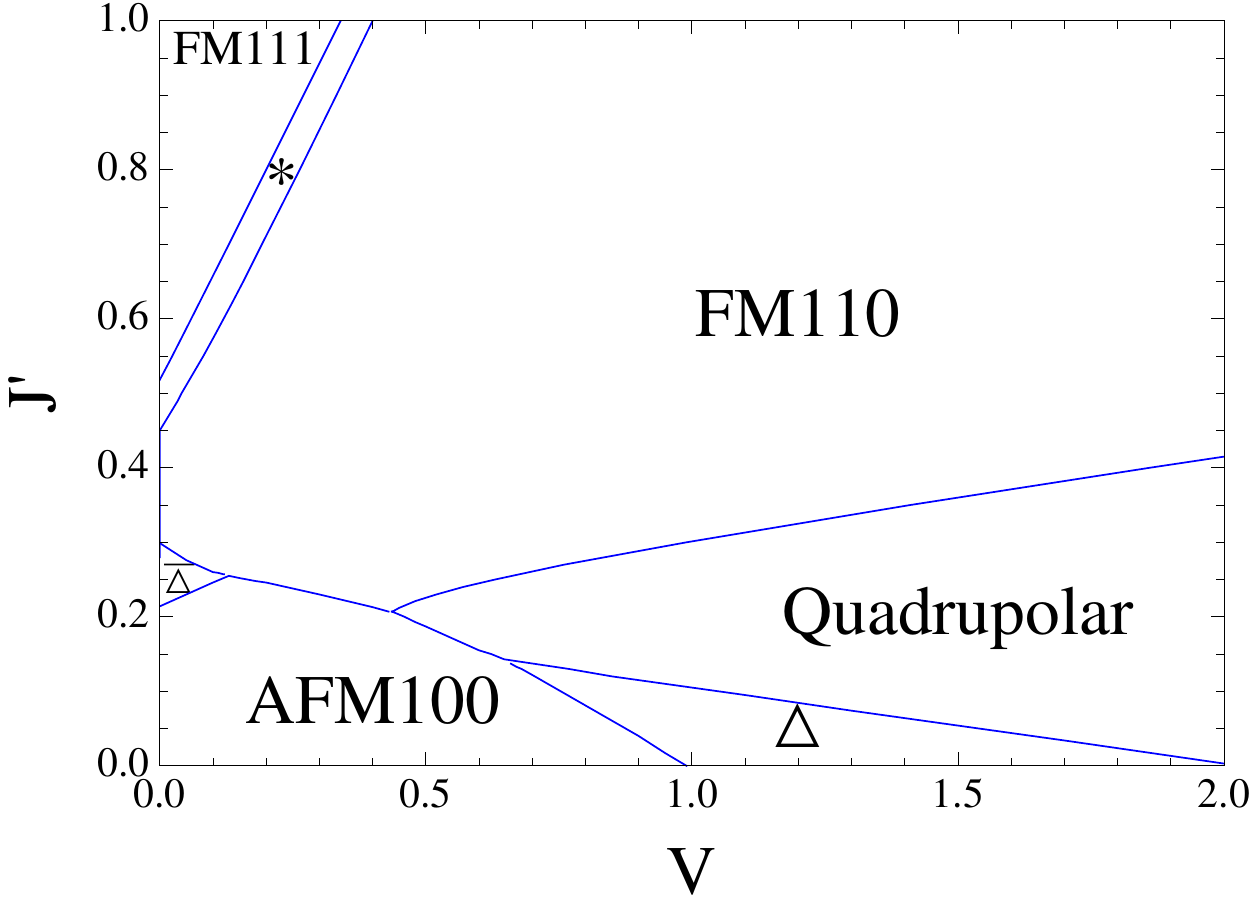}
\caption{ (Color online) Ground state phase diagram of the model
  Hamiltonian in Eq.~\eqref{eq:ham}.  Here $V$ is the electric
  quadrupole interaction, and $J'$ is the nearest-neighbor ferromagnetic
  exchange between orthogonal orbitals; both are normalized to a unit
  value, $J=1$, for the nearest-neighbor antiferromagnetic coupling.
  The names for the phases are defined in Table~\ref{tab:phases}, and in
  Sec.~\ref{sec:sec3}.}
\label{fig:pd1}
\end{figure}

The results of a mean-field analysis are summarized in
Fig.~\ref{fig:pd1} and Table~\ref{tab:phases}.  There are seven total
ground state phases which appear, including notably a broad region of
time-reversal invariant but quadrupolar ordered (spin nematic) ground
state with a two-sublattice structure of the orbital configuration.
This is described by the quadrupole tensor operators,
\begin{eqnarray}
  \label{eq:7}
  Q_i^{3z^2} & = & [2(j_{i}^z)^2-  (j_i^x)^2-(j_i^y)^2]/\sqrt{3} , \\
  Q_i^{x^2-y^2} & = & (j_i^x)^2-(j_i^y)^2.
\end{eqnarray}
In the quadrupolar phase,$\langle Q_i^{x^2-y^2}\rangle = \pm q'$
alternates sign on the two sublattices (see Eqs.~\eqref{eq:8}).  In
this phase, the time reversal symmetry is {\sl unbroken} and there is
no magnetic dipolar and octupolar order.  Such a {\sl local} spin
nematic ground state is particular to integer spin systems and
prohibited for half-integer spins.  A similar spin nematic ground
state has also been proposed theoretically for a spin $S=1$ material
NiGa$_2$S$_4$,\cite{PhysRevB.79.214436,quadrupolar2006,JPSJ.75.083701,PhysRevB.78.220403,PhysRevB.74.092406,Nakatsuji09092005}
but an experimental confirmation of this phase in NiGa$_2$S$_4$ is
still lacking.  Apart from the quadrupolar ordered phase, the other
ground states are magnetic, and comprise both ferromagnetic and
antiferromagnetic states with enlarged unit cells up to quadruple the
size of the ideal one.  The extension of the mean field analysis to
$T>0$ is described in Fig.~\ref{fig:pd2}. Amongst various finite
temperature phases, the quadrupolar ordered one is prominent,
providing more opportunity for its experimental discovery in real
material.

The remainder of the paper is organized as follows.  In
Sec.~\ref{sec:sec2}, we first explain the on-site spin and orbital
physics which leads to an effective $j=2$ local moments on each B'
magnetic ion.  As we did in previous work,\cite{PhysRevB.82.174440} we
introduce a microscopic Hamiltonian which includes three interactions:
nearest neighbor (NN) antiferromagnetic (AFM) exchange, $J$, NN
ferromagnetic (FM) exchange, $J'$, and electric quadrupolar interaction,
$V$. Due to strong SOC, we project these interactions, which are written
in the separate spin and orbital spaces, down to the $j=2$
manifold. This leads to a spatially anisotropic and very
non-Heisenberg-like Hamiltonian, which contains many terms beyond the
usual quadratic exchange.
\begin{table}
  \centering
  \begin{tabular}{ccccc}
    Phase & $\vec{m}$ & $\vec{n}$ & magnetic & quadrupolar \\
    & & & unit cell & unit cell \\
    \hline
    AFM100 & 0 & [100] & 2 & 1 \\
    FM111 & [111] & 0 & 1 & 1 \\
    FM110 & [110] & [1$\overline{1}$0] & 2 & 2\\
    * & [11x] & [1$\overline{1}$0]  & 2 & 2 \\
    $\Delta$ & 0 & [xy0] & 4 & 2 \\
    $\overline{\Delta}$ & 0 & [xy0] & 4 & 1 \\
    quadrupolar & 0 & 0 & 2 & 2 
  \end{tabular}
  \caption{Phases in mean field theory.  The first column gives the
    name of the phase, as shown in Fig.~\ref{fig:pd1}.  In the second column, $\vec{m}$ denotes the
    direction (up to symmetries) of the uniform magnetization; a ``0''
    is shown if this is vanishing.  Similarly, in the third column,
    $\vec{n}$ denotes the direction of the staggered magnetization.  For
    the quadrupolar phase, in which
    ``0'' appears in both the second and third columns, time-reversal
    symmetry is unbroken.    The fourth column specifies the number of
    sites in the magnetic unit cell.  The fifth column gives the number
    of site in the unit cell for time-reversal invariant observables,
    i.e. the quadrupolar tensor. }
  \label{tab:phases}
\end{table}
In Sec.~\ref{sec:sec3}, we find the mean field ground state phase
diagram of the model Hamiltonian and analyze the properties of each
ground state.   Some insight into the general results is given by
considering some simple limits, including both strong easy plane and
easy axis anisotropy, and an orbital interaction only model
(Sec.~\ref{sec:sec31}).  We find that, besides a uniform orbitally
ordered  Finally, in Sec.~\ref{sec:sec32},
we carry out a mean field study for the complete model Hamiltonian and
find the zero temperature ground state phase diagram which is depicted
in Fig.~\ref{fig:pd1}.  

In Sec.~\ref{sec:sec4}, we extend the mean field theory to $T>0$, and
identify the structure of the magnetic order for each phase. We find ten
different finite temperature magnetization processes which correspond to
the ten overlapping regions between the ground state phases and the
shaded areas in Fig.~\ref{fig:pd2}. We also discuss the different finite
temperature phase transitions associated with magnetic dipolar and
quadrupolar orders.  Transition temperatures are extracted from Landau
theory. In a broad region of the parameter space, we find the
quadrupolar phase occurs at the intermediate temperatures between the
high temperature paramagnetic phase and various low temperature magnetic
ordered phases.

Finally in Sec.~\ref{sec:sec5}, we compare our theoretical predictions
with current experimental findings and suggest further directions for
theory and experiment.

\section{Model}
\label{sec:sec2}

\subsection{Spin-orbit interaction and electron orbitals}
\label{sec:sec21}

The magnetic ions B' (Re$^{5+}$, Os$^{6+}$) in the relevant ordered
double perovskites (Ba$_2$CaOsO$_6$, La$_2$LiReO$_6$,
Ba$_2$YReO$_6$)\cite{**} all have a $4d^2$ or $5d^2$ electron
configuration with two electrons on the triply degenerate $t_{2g}$
multiplets.  Because of the electron interaction, the local spin orbital
state is quite different from the case of $d^1$ electron configuration
where the single electron state is enough to describe the local physics.
Considering the dominance of the crystal field splitting over the SOC,
we now fill the three $t_{2g}$ orbitals with these two electrons before
including the effect of SOC.  To respect the first Hund's rule, the
total spin for the two electrons is $S=1$.  For the orbital sector,
there are three degenerate antisymmetric two-electron states,
\begin{eqnarray}
|\text{X}\rangle &=& \frac{1}{\sqrt{2}} (  |xy\rangle_1 |xz\rangle_2 -|xy\rangle_2 | xz \rangle_1   ) \\
|\text{Y}\rangle &=& \frac{1}{\sqrt{2}} (  |xy\rangle_1 |yz\rangle_2 -|xy\rangle_2 | yz \rangle_1   ) \\
|\text{Z}\rangle &=& \frac{1}{\sqrt{2}} (  |xz\rangle_1 |yz\rangle_2 -|xz\rangle_2 | yz \rangle_1   ) 
\;,
\label{eq:onsite_orb}
\end{eqnarray} 
in which, the subindex (``1'' and ``2'') labels the electron. 
Therefore, there are totally nine-fold spin-orbital degeneracies. The presence of SOC will lift some of 
the degeneracies. Following the spirit of degenerate perturbation theory, we project the SOC onto the 
triplet subspace spanned Eq.~\eqref{eq:onsite_orb},
\begin{equation}
{\mathcal H}_{\text{so}} = -\lambda\; {\bf l} \cdot {\bf S} 
\;,
\label{eq:soi}
\end{equation}
in which, the total angular momentum quantum number of these operators
are $l=1, S=1$.  The effective orbital angular momentum ${\bf l}$ comes
from the projection of the {\sl total} orbital angular momentum $ {\bf
  L} \equiv {\bf L}_1 +{\bf L}_2 $ onto the tripets in
Eq.~\eqref{eq:onsite_orb},
\begin{equation}
{\mathcal P}_o \ {\bf L}  \ {\mathcal P}_o = - {\bf l}
\;.
\end{equation} 
Here ${\mathcal P}_o \equiv \sum_{\text{A}=\text{X,Y,Z}}
|\text{A}\rangle \langle \text{A}|$ is the projection operator to the
triplet orbital subspace.

The reduced SOC in Eq.~\eqref{eq:soi} favors a local $j=2$ (${\bf j} =
{\bf l} + {\bf S}$ ) over other higher energy states $j=0,1$ by an
energy separation $\mathcal{O}(\lambda) $.  In the materials we are
considering, the SOC $\lambda$ is a very large energy scale (some
fraction of an eV).

In general, cubic symmetry allows the presence of an on-site cubic
anisotropy term, $(j^x)^4 + (j^y)^4 + (j^z)^4$, which lifts the
degeneracy of the five $j=2$ states.  However, we expect this splitting
to be rather small, and provided it is smaller than the typical exchange
coupling between spins, the $j=2$ description should be a good
approximation. Microscopically the cubic anisotropy comes from the 4th
order effect of the SOC and pair hopping (between different orbitals on
the same ion) terms $J_p$ which excite the electrons into the $e_g$
orbitals.  The magnitude of the cubic anisotropy should be of $\sim
{\mathcal O}(\lambda^4/\Delta^3, J_p^4/\Delta^3)$ (with $\Delta$ the
crystal field splitting between $e_g$ and $t_{2g}$ levels).  This is
certainly a much smaller energy scale compared to SOC, and likely small
compared to exchange.  In any case, we will neglect it in the following.

In the strong SOC limit, every local operator should be projected onto
the local subspace spanned by five $j=2$ states. In particular,
\begin{eqnarray}
{\mathcal P}_2 \ {\bf S}  \ {\mathcal P}_2 &=& \frac{1}{2 } {\bf j}\\ 
{\mathcal P}_2 \ {\bf l}  \ {\mathcal P}_2 &=& \frac{1}{2 } {\bf j} 
\;.
\end{eqnarray}
Here ${\mathcal P}_2$ is the projection operator into the local $j=2$
states.  In addition, one can find the local magnetic moment is given by
\begin{equation}
{\bf M} = {\mathcal P}_2 \ ( 2 {\bf S} - {\bf l}  ) \ {\mathcal P}_2 = \frac{1}{2} {\bf j }
\;,
\end{equation}
hence, the magnitude of the local magnetic moment is found to be $\sqrt{6}/2 \mu_{\text B} \approx 1.25 \mu_{\text B}$.

\subsection{Exchange interactions and electric quadrupolar interaction}

In this subsection, we introduce the interactions between the local
moments. From previous work,\cite{PhysRevB.82.174440} we will need to
consider the nearest neighbor (NN) antiferromagnetic (AFM) exchange, NN
FM exchange and NN electric quadrupolar interactions, and these
interactions are highly anisotropic in both the position space and spin
space. For example, in the XY plane, only electrons on $xy$ orbital can
virtually transfer from one site to another via the intermediate oxygen
$p$ orbitals. Thus, one finds that the NN AFM exchange is written as
\begin{equation}
{\mathcal H } _{\text{AFM}}^{\text{XY}}   = J \sum _{\langle ij \rangle \in \text{XY}} \big[ {\bf S}_{i,xy} \cdot {\bf S}_{j,xy} - \frac{1}{4} n_{i,xy} n_{j,xy} \big]
\;,
\end{equation}
where the sum is over nearest neighbor sites in the XY planes, and the
correponding terms for YZ and XZ planes can be obtained by the obvious
cubic permutation. One should note that the operators ${\bf S}_{i,xy} $
and $n_{i,xy}$ denote the electron spin residing on the single-electron
$xy$ orbital and orbital occupation number for the single-electron $xy$
orbital at site $i$, respectively.  To connect these single electron
operators to the two-electron operator which acts on the two-electron
orbitals in Eq.~\eqref{eq:onsite_orb}, we have the following relations
\begin{eqnarray}
n_{i,xy}  & = & n_{i, \text{X}} + n_{i,\text{Y}} = (l_i^z)^2\\
{\bf S}_{i,xy} &=& \frac{{\bf S}_i}{2} (n_{i,\text{X}} + n_{i,\text{Y}}) = \frac{ {\bf S}_i  }{2}  (l_i^z)^2
\;.
\label{eq:mapping}
\end{eqnarray}
Here $n_{i,\text{X}}$ (or $n_{i,\text{Y}}$) denotes the occupation number for $| \text{X}\rangle$ 
(or $|\text{Y} \rangle$) of the two-electron orbital states at site $i$, and ${\bf S}_i$ is the total 
spin $S=1$ for the two electrons. The physical meaning of Eq.~\eqref{eq:mapping} is apparent. 
The electron occupation number on the single-electron orbital $xy$ can be nonvanishing only when the 
two-electron orbital state $|\text{X}\rangle$ or $| \text{Y} \rangle$ is occupied by the two electrons.

Throughout this paper, we use the subindices ($i,xy$) to denote the site and single electron orbitals,  
subindex $\text{X}$ to denote the two-electron orbitals, superindex ($\mu =x,y,z$) to denote the spin 
component, and capital letters (XY, XZ, YZ) to denote the planes. With these definitions, 
we note the double occupancy condition at each site, which defines the Mott insulating phase,
becomes,
\begin{equation}
n_{i,xy} + n_{i,xz} + n_{i,yz} =2
\;,
\end{equation}
in terms of the two-electron operators, the above equation is equivalent to
\begin{equation}
n_{i,\text{X}} + n_{i,\text{Y}} + n_{i,\text{Z}} =1
\;.
\label{eq:orb_constraint}
\end{equation}
Moreover, from Eq.~\eqref{eq:mapping}, orbitally-resolved spins satisfy
\begin{equation}
{\bf S}_{i,xy}+{\bf S}_{i,xz}+{\bf S}_{i,yz} = {\bf S}_i 
\;.
\end{equation}

The second interaction to include is the NN FM exchange interaction.  FM
exchange comes about when the orthogonal $p$ orbitals on a single oxygen
ion are involved in the exchange path. Directly using the results from
Ref.~\onlinecite{PhysRevB.82.174440} and the relation in
Eq.~\eqref{eq:mapping}, one can immediately write down this
interaction. Again, for two sites $i,j$ in the XY plane, this FM exhange
is given as
\begin{eqnarray}
{\mathcal H}_{\text{FM},ij}^{\text{XY}} &=& -J' \big[{\bf S}_{i,xy} \cdot ( {\bf S}_{j,yz} 
                     + {\bf S}_{j,xz} ) + \langle i \leftrightarrow j \rangle \big] \nonumber \\
            && + \frac{3J'}{2} n_{i,xy} n_{j,xy}
\nonumber \\
&=&  - \frac{J'}{4} \big[ {\bf S}_i \cdot {\bf S}_j (l_i^z)^2 \big( (l_j^x)^2 + (l_j^y)^2 \big) 
         + \langle i \leftrightarrow j \rangle \big]   \nonumber \\
&  &  + \frac{3J'}{4} (l_i^z)^2(l_j^z)^2
\;.
\end{eqnarray}

The third interaction to include is the electric quadrupolar
interaction.  This is obtained by evaluating the Coulomb interaction
in different orbital occupations.  As the AFM and FM exchange, we
also take results from previous work to write down this interaction.  In
XY plane, we obtain the electric quadrupolar interaction as
\begin{eqnarray}
{\mathcal H} _{\text{quad},ij}^{\text{XY}} &=&
-\frac{4V}{3} (n_{i,xz} - n_{i,yz}) (n_{j,xz} - n_{j,yz}) \nonumber \\
&& + \frac{9V}{4} n_{i,xy} n_{j,xy} \nonumber \\
&=& - \frac{4V}{3} \big(  (l_i^y)^2 - (l_i^x)^2  \big) \big(  (l_j^y)^2 - (l_j^x)^2  \big) \nonumber \\
&& + \frac{9V}{4} (l_i^z)^2(l_j^z)^2\label{eq:1}
\end{eqnarray}

The minimal Hamiltonian for the cubic system contains all three of these interactions in addition to the onsite SOC, 
\begin{equation}
{\mathcal H} = {\mathcal H}_{\text{AFM}} + {\mathcal H}_{\text{FM}} + {\mathcal H}_{\text{quad}} + {\mathcal H}_{\text{so}}
\;.
\end{equation}

Since we are interested in the limit of strong SOI, we have to project the minimal Hamiltonian ${\mathcal H}$ 
onto the five $j=2$ states at every site. As an illustration, we write down the projection for ${\bf S}_{i,xy}$ and $n_{i,xy}$,
\begin{eqnarray}
\tilde{S}_{i,xy}^x &=& \frac{1}{12} j^x_i + \frac{1}{12} (j^x_i)^3 ,\\
\tilde{S}_{i,xy}^y &=& \frac{1}{12} j^x_i + \frac{1}{12} (j^x_i)^3 , \\
\tilde{S}_{i,xy}^z &=& -\frac{1}{12} j^x_i + \frac{1}{12} (j^x_i)^3 ,  \\
\tilde{n}_{i,xy} &=&\frac{1}{3}+ \frac{1}{6} (j^z_i)^2 ,
\label{eq:so_proj}
\end{eqnarray}
in which, $\tilde{O} = {\mathcal P}_2 O {\mathcal P}_2$. After the projection, the minimal Hamiltonian that we 
will study in this paper is
\begin{equation}
\tilde{\mathcal H} = \tilde{\mathcal H}_{\text{AFM}} +\tilde{\mathcal H}_{\text{FM}} + \tilde{\mathcal H}_{\text{quad}}.
\label{eq:ham}
\end{equation}

\section{Mean-field ground states}
\label{sec:sec3}

In this section, we study the zero temperature phase diagram of the
model Hamiltonian in Eq.~\eqref{eq:ham}.  Ultimately, in
Sec.~\ref{sec:sec32}, we will do this by mean field theory, or
equivalently, variationally searching for direct product states which
minimize the expectation of $\tilde{\mathcal H}$.  Before reporting
these results, however, we discuss some simple limits in which the
behavior can be understood more intuitively.  First, in
Sec.~\ref{sec:uniaxial-anisotropy}, we impose a strong single-ion
uniaxial anisotropy, which removes the orbital degeneracy renders the
problem trivially soluble with singlet, ferromagnetic and
antiferromagnetic ground states.  Second, we consider a pure orbital
model, in which only the electric quadrupole interaction $V$ is
included.  This gives the two sublattice quadrupolar
state described in the introduction.  Finally, in Sec.~\ref{sec:sec32} we report
the results of a full mean-field calculation including all couplings and
making no further approximations.

\subsection{Uniaxial anisotropy}
\label{sec:uniaxial-anisotropy}

As we did in Ref.~\onlinecite{PhysRevB.82.174440}, we first consider the
ground state of this Hamiltonian in the presence of strong easy-plane or
easy-axis anisotropies. The strong easy-plane anisotropy (on XY plane)
is a trivial limit and is modeled by $\sum_i D (j_i^z)^2$ with a
positive $D$. When $D$ is quite large (compared to exchange coupling and
electric quadrupolar interaction), the spin state on every site is
pinned to $|j^z =0\rangle$, which is a rather trivial uniform state with
an ordering wavevector ${\bf p} = {\bf 0}$.  The strong easy-axis
anisotropy (along $z$ direction) is less trivial and is modeled by the
same Hamiltonian but with a negative $D$. Large $|D|$ favors either
$|j^z =2\rangle$ or $|j^z = -2\rangle$ to be occupied.  After projecting
the Hamiltonian in Eq.~\eqref{eq:ham} onto this two states, the electric
quadrupolar interaction is completely quenched and the resulting
effective Hamiltonian is a trivial Ising Hamiltonian.  One can readily
find that, when $J' \geq 5J/38$ the ground state is a ferromagnetic
state with an ordering wavevector ${\bf p} ={\bf 0}$, and when $J' \leq
5J/38$ the ground state is an antiferromagnetic state with an ordering
wavevector ${\bf p} =2\pi(100)$ or $2\pi(010)$. One may postulate from
these anisotropic case that, the ground state for the actual cubic
Hamiltonian may either have a uniform state (${\bf p} = {\bf 0}$) or a
two-sublattice state (${\bf p}=2\pi(001)$ and equivalent
wavevectors). As we will see in the following sections, this guess
is correct for a large portion of the parameter space, but we also find
some interesting exceptions. 

\subsection{The orbital Hamiltonian}
\label{sec:sec31}

To understand the nature of the non-magnetic quadrupolar ground state
of the model, it is sufficient to consider only the electric
quadrupole interaction $V$, given in Eq.~\eqref{eq:1}.  Note that it
involves only the three operators
$\tilde{n}_{i,yz}$, $\tilde{n}_{i,xz}$ and $\tilde{n}_{i,xy}$ on each
site.   Since these are all time-reversal invariant, it is apparent
that they do not span the full space of $j=2$ operators.  Thus there
must be additional constants of the motion, and the Hamiltonian can be
separated into sectors corresponding to different irreducible
representations (irreps) of the algebra of these operators.  Indeed, one can
show that the single-site Hilbert space decomposes into one
two-dimensional irrep and three one-dimensional irreps.  The two
dimensional irrep is spanned by the two states
\begin{eqnarray}
  \label{eq:2}
  |u\rangle & = & \frac{1}{\sqrt{2}}\left( |j^z=2\rangle +
    |j^z=-2\rangle\right), \\
  |d\rangle & = & |j^z=0\rangle.
\end{eqnarray}
In this subspace, the orbital operators become
\begin{eqnarray}
  \label{eq:3}
  \tilde{n}_{i,yz} & = & \frac{2}{3} - \frac{1}{6} \sigma^z_i +
\frac{1}{2\sqrt{3}}\sigma_{i}^x, \\
  \tilde{n}_{i,xz} & = & \frac{2}{3} - \frac{1}{6} \sigma^z_i -
\frac{1}{2\sqrt{3}}\sigma_{i}^x, \\
  \tilde{n}_{i,xy} & = & \frac{2}{3} +\frac{1}{3} \sigma^z_i ,
\end{eqnarray}
where $\vec\sigma_i$ are the Pauli matrices acting in the
$|u\rangle,|d\rangle$ space.
In the first of the three one dimensional irreps we have 
\begin{equation}
  \label{eq:4}
  \begin{pmatrix} \tilde{n}_{i,yz} \\ \tilde{n}_{i,xz} \\
    \tilde{n}_{i,xy} \end{pmatrix} =  \begin{pmatrix} 1/2 \\
    1/2 \\ 1\end{pmatrix}.
\end{equation}
The other two one dimensional irreps may be obtained by permuting
these values.

The ground state must consist of a single irrep on each site.  Which
irrep occurs should be determined by minimization of the
energy. Consider the case in which each site has a one-dimensional
irrep.  Then we can specify the state of a site by a ``Potts''-like
variable $s_i=1,2,3$ specifying which of the three orbital numbers
equals 1, i.e.
\begin{equation}
  \label{eq:1}
  \tilde{n}_{i,a} = \frac{1}{2} + \frac{1}{2}\delta_{s_i,a},
\end{equation}
with $a=1,2,3$ corresponding to $a=yz,xz,xy$, respectively.  Then the
Hamiltonian for a bond in the XY plane becomes
\begin{eqnarray}
  \label{eq:5}
  \left.\tilde{\mathcal H}^{XY}_{ij} \right|_{\textrm{1d irreps}}& =  & -\frac{V}{3}
  \left(\delta_{s_i,1}-\delta_{s_i,2}\right)
  \left(\delta_{s_j,1}-\delta_{s_j,2}\right) \nonumber \\
  && +
  \frac{9V}{16}\left(1+\delta_{s_i,3}\right) \left(1+\delta_{s_j,3}\right),
\end{eqnarray}
with YZ and XZ plane bond interactions obtained by permutations.  In
this case the Hamiltonian is purely classical and thus the ground state
can be exactly found by minimization.  We find that a ferromagnetic
ground state is preferred, with constant $s_i$ (there are thus three
such degenerate ground states), which has an average energy of
$65V/72\approx 0.903V$ per bond.

The other natural choice to consider is when the two-dimensional irrep
is chosen on each site.  In this case the Hamiltonian can be written as
\begin{equation}
  \label{eq:6}
   \left.\tilde{\mathcal H}^{XY}_{ij} \right|_{\textrm{2d irrep}}=
   V\left[1 - \frac{25}{36} \sigma_i^x \sigma_j^x +
     \frac{1}{4}\vec{\sigma}_i \cdot \vec{\sigma}_j\right],
\end{equation}
where the dot product involves only the x and z components of the
Pauli matrices, and we have dropped a term linear in the Pauli
matrices which cancels when all three types of bonds are added.  In
this case the interactions for bonds in the other two planes are
obtained by the rotations $\sigma_i^x \rightarrow
-\frac{1}{2}\sigma_i^x \pm \frac{\sqrt{3}}{2} \sigma_i^z$, with the
upper (lower) sign chosen for the XZ (YZ) plane.  The dot product is
unchanged by this rotation.

The Hamiltonian in this subspace is a type of Kugel-Khomskii model,
similar to that studied in models of $e_g$ orbitals.  It is fully
quantum, and thus cannot be solved exactly.  However, within mean field
theory we find the variational ground state is simply the
anti-ferromagnetic product state with $\sigma_i^x=(-1)^z$, i.e. with
alternating sign on adjacent XY planes.  In this
state the expectation of the Hamiltonian can be taken, and the energy is
found to be $E_v= 173V/216 \approx 0.801V$ per bond.  Note that this is
lower than the energy found for the one-dimensional irreps.    This is an upper
(variational) bound on the ground state energy in this sector, so we
indeed expect the ground state to be in the two-dimensional irrep.  

Physically, the mean-field state describes an orbitally ordered phase
with a two sublattice structure with single-ion wavefunctions
\begin{eqnarray}
|\text {A} \rangle &=& \frac{1}{2} | j^z = 2\rangle + \frac{1}{\sqrt{2}} | j^z=0 \rangle + \frac{1}{2} | j^z=-2 \rangle ,
\\
|\text{B} \rangle &=& \frac{1}{2} | j^z = 2\rangle - \frac{1}{\sqrt{2}} | j^z=0 \rangle + \frac{1}{2} | j^z=-2 \rangle ,
\end{eqnarray}
where A, B label the two different sublattices (planes with even/odd
$z$). Remarkably, these states are {\sl invariant} under time
reversal. Since time reversal symmetry is unbroken, there is no
magnetic order,
\begin{equation}
\langle {\bf j}_i \rangle = 0.
\end{equation}
Therefore, this phase is a magnetic quadrupolar phase (or spin nematic phase) 
and the orbital configuration is given by
\begin{eqnarray}
\langle \tilde{\bf n}_{\text{A}} \rangle &=& ( \frac{2}{3} + \frac{\sqrt{3}}{6}, \frac{2}{3} - \frac{\sqrt{3}}{6}, \frac{2}{3} ) \\
\langle \tilde{\bf n}_{\text{B}} \rangle &=& ( \frac{2}{3} - \frac{\sqrt{3}}{6}, \frac{2}{3}  + \frac{\sqrt{3}}{6}, \frac{2}{3} ). 
\label{eq:orb_phaseT}
\end{eqnarray}
Here, we defined $\tilde{\bf n} \equiv (\tilde{n}_{yz},\tilde{n}_{xz},\tilde{n}_{xy})$ for convenience.

\subsection{The full cubic Hamiltonian}
\label{sec:sec32}

Although both the anisotropic limit and pure orbital interaction
support a two-sublattice ground state, it is still questionable that
the cubic Hamiltonian will also behave likewise.  In this section we
report the results of a systematic investigation of the mean field
ground states of the full Hamiltonian, allowing for large unit cells
(we considered cells of up to 4 sites).  We made no further
assumptions and variationally minimized the energy with respect to an
arbitrary wavefunction on every site of the unit cell.  Finally we
verified that each mean field ground state is stable within linear
flavor wave theory.\cite{PhysRevB.82.174440} The mean field phase
diagram is depicted in Fig.~\ref{fig:pd1}, and the key features of
each phase is listed in Table~\ref{tab:phases}.  Within the flavor
wave theory, all seven phases exhibit an energy gap.  Next we describe
each of the seven phases.

\subsubsection{Antiferromagnetic (AFM100) state}
\label{sec:sec321}

For small $J'/J$ and $V/J$, the ground state is a typical antiferromagnetic phase with an
ordering wavevector ${\bf p} = 2\pi (001)$.  States with the
equivalent momenta ${\bf p}=2\pi(100),2\pi(010)$ are of course degenerate. To
be specific, we will take ${\bf p} = 2\pi (001)$ for all the
two-sublattice phases in the following.  The variational
mean field ground state wavefunction on the A and B sublattices has
the following form: 
\begin{eqnarray}
| \text{A} \rangle &=& \frac{x}{\sqrt{2}}( |j^z=2 \rangle + | j^z=-2 \rangle) + \sqrt{\frac{1}{2}-x^2} | j^z = 0\rangle
\nonumber \\
&+&\frac{1}{2} ( |j^z=1\rangle + | j^z=-1 \rangle) 
\\
| \text{B} \rangle &=& \frac{x}{\sqrt{2}}( |j^z=2 \rangle + | j^z=-2 \rangle) + \sqrt{\frac{1}{2}-x^2}  | j^z = 0\rangle
\nonumber \\
&-& \frac{1}{2} ( |j^z=1\rangle + | j^z=-1 \rangle),
\label{eq:AFM_fct}
\end{eqnarray}
in which $x$ is a real parameter, which is found by minimizing the
variatonal energy. Since under time-reversal, $|j^z=m\rangle
\rightarrow (-1)^m |j^z=m\rangle$, these two states transform into one
other under time reversal. Therefore, the magnetic dipolar and
octupolar orders are anti-parallel on two sublattices.  Because of the
intrinsic strong SOC, the local moments are aligned by
crystalline anisotropy, and in this state orient along the [100] (or
equivalently, [010]) axis.   A more precise description of the symmetry breaking
of the phase is given by introducing the magnetic dipole and
quadrupole moment operators.  The magnetic dipole
moment is antiferromagnetically ordered,
\begin{equation}
\langle {\bf j}_{\text{A} / \text{B}}\rangle = \pm m (1,0,0), 
\end{equation}
with $m=\sqrt{2} x + \sqrt {3(1-2x^2)}$. The magnetic quadrupole
tensor, however, is uniformly ordered,
\begin{eqnarray}
\langle Q_{i}^{3z^2} \rangle & \equiv &  2\sqrt{3} (3 \langle \tilde{n}_{i, xy} \rangle-2) = q
\\
\langle Q_{i}^{x^2 - y^2} \rangle & \equiv & 6 \langle \tilde{n}_{i,yz} 
-\tilde{n}_{i,xz}  \rangle = q'
\end{eqnarray}
with $q= 2\sqrt{3} (2x^2 -3/4)$ and $q'= 2 \sqrt{3 (1-2x^2) } + 3/2 $.

The uniform quadrupolar order (or orbital configuration) can be
understood to arise from the large NN AFM exchange $J$ which favors
time reversal pairs on two sublattices. As the ferromagnetic exchange
and electric quadrupolar interaction increase, the orbital-orbital
interaction will become important and the uniform orbital structure
will break down.

\subsubsection{Uniform ferromagnetic (FM111) state }
\label{sec:sec322}

With large $J'/J$ and small $V/J$, the ferromagnetic exchange
dominates and favors a uniform ground state with the spin polarization
aligned with [111] or other equivalent lattice directions.  The
mean-field ground state of this phase is a fully polarized
spin eigenstate with quantization axis along [111], so that
\begin{equation}
\langle {\bf j}_i \rangle = \frac{m}{\sqrt{3}} (1,1,1),
\end{equation}
with $m=2$. 
And the three orbitals are equally populated,
\begin{equation}
\langle \tilde{\bf n}_i \rangle = (\frac{2}{3}, \frac{2}{3}, \frac{2}{3}). 
\end{equation}
and the magnetic quadrupolar orders vanish,
\begin{equation}
\langle Q_i^{3z^2} \rangle = \langle Q_i^{x^2 - y^2} \rangle = 0.
\end{equation}

\subsubsection{Two-sublattice ferromagnetic (FM110) state}
\label{sec:sec323}

With large $J'/J$ and $V/J$, we obtain a FM110 state which is the same
phase proposed in Ref.~\onlinecite{PhysRevB.82.174440}.  This state
can be considered as a compromise between the tendencies of $J$ and
$J'$ to order the moments along the [100] and [111] axes.  The
competition between these two effects allows the orbital interaction
to stabilize a coexisting quadrupolar order.  The result is that the
orbital configuration has a two-sublattice structure, and the magnetic
moments on the two sublattices are neither antiparallel or
parallel. The ground state wavefunction of this phase is
parametrized by two complex numbers, $x_1$ and $x_2$,
\begin{eqnarray}
| \text{A} \rangle &=& \frac{1}{\sqrt{2}}( x_2 |j^z=2 \rangle + \overline{x_2} | j^z=-2 \rangle) 
+x_0| j^z = 0\rangle
\nonumber \\
&+&\frac{1}{\sqrt{2}} ( x_1 |j^z=1\rangle + \overline{x_1} | j^z=-1 \rangle) 
\\
| \text{B} \rangle &=& \frac{1}{\sqrt{2}}( -\overline{x_2}|j^z=2 \rangle - x_2 | j^z=-2 \rangle) 
+x_0| j^z = 0\rangle
\nonumber \\
&+&\frac{1}{\sqrt{2}} ( -i \overline{x_1} |j^z=1\rangle +ix_1| j^z=-1 \rangle) 
\end{eqnarray}
where, $x_0 = \sqrt{1-|x_1|^2-|x_2|^2} $ and $\overline{x_{1,2}}$ is
complex conjugation of $x_{1,2}$.  From the wavefunction, we find that
magnetic dipole and quadrupole moments have the following form:
\begin{eqnarray}
\langle {\bf j}_{\text{A} / \text{B} } \rangle & = & m \frac{1}{\sqrt{2}}(1,1,0) \pm m' \frac{1}{\sqrt{2}}(1,-1,0) 
\\
\langle  Q_{\text{A} / \text{B} }^{3z^2}  \rangle & = & q
\\
\langle Q_{\text{A} / \text{B}} ^{x^2- y^2 } \rangle & = & \pm q'.
\label{eq:FM110_order}
\end{eqnarray}
Here, the actual expression of $m, m', q, q' $ in terms of $x_1$ and
$x_2$ is quite involved and not important for the purpose of
presentation.  So we will not write them out explicitly. Similar
omissions are made in later sections.

From Eq.~\eqref{eq:FM110_order}, the total magnetic moment is along
[110] direction, and the staggered magnetic dipole moment is
perpendicular to the total magnetic dipole moment and along
[$1\bar{1}0$] direction.  The orbital configurations of the two
sublattices are similar to that in the quadrupolar phase (see
Eq.~\eqref{eq:orb_phaseT}).

\subsubsection{Intermediate ferromagnetic (``$\ast$'') state}
\label{sec:sec324}

Between the FM111 and FM110 phases, we find an intermediate state,
which we denote ``$\ast$''.  This state also has a two-sublattice
structure with the ordering wavevector ${\bf p} = 2\pi (001)$.  We
find the ground state wavefunction of this phase is parametrized by
four complex numbers, $x_1, x_2, x_{-1}$ and $x_{-2}$,
\begin{eqnarray}
|\text{A} \rangle &=& x_2 | j^z=2 \rangle+  x_1 | j^z =1 \rangle + x_0 | j^z=0\rangle
\nonumber \\
&+& x_{-1}| j^z=-1\rangle+ x_{-2} | j^z =-2 \rangle  \\
|\text{B}  \rangle &=& -\overline{ x_2}| j^z=2 \rangle -i  \overline{x_1} | j^z =1 \rangle + x_0  | j^z=0\rangle
\nonumber \\
&+& i \overline{x_{-1}}| j^z=-1\rangle - \overline{ x_{-2}}
| j^z =-2 \rangle
\label{eq:star_fct}
\end{eqnarray}
with $x_0 = \sqrt{1- |x_1|^2-|x_2|^2-|x_{-1}|^2-|x_{-2}|^2}$.  The
magnetic order in this phase interpolates between that of the FM111
and FM110 phases.   The uniform magnetization orients along an axis
between the [111] and [110] directions.   Like in FM110
phase, the staggered magnetic dipole moment in phase ``$\ast$'' is 
along [$1\bar{1}0$] direction.  The orbital configuration has
the same two-sublattice structure as in FM110 phase.  The corresponding
order parameters are
\begin{eqnarray}
\langle {\bf j}_{\text{A} / \text{B} }\rangle & = & (m_1,m_1,m_2) \pm m' \frac{1}{\sqrt{2}}(1,-1,0) ,
\\
\langle  Q_{\text{A} / \text{B} }^{3z^2}  \rangle & = & q,
\\
\langle Q_{\text{A} / \text{B}} ^{x^2- y^2 } \rangle & = & \pm q'.
\end{eqnarray}

\subsubsection{Four-sublattice antiferromagnetic (``$\Delta$'') state}
\label{sec:sec325}

In the region of a small $J'/J$ and an intermediate $V/J$, the FM
exchange interaction between time-reversally odd moments has negligible
effec, while the AFM exchange and electric quadrupolar coupling are
somewhat balanced. The keen competition between these two interactions
induces an interesting intermediate phase: a four-sublattice
antiferromagnetic phase which we denote``$\Delta$''.  The magnetic unit
cell is the elementary tetrahedron of the fcc lattice.  Similar to FM110
phase, the ground state wavefunction is found to be parametrized by two
complex number, $x_1$ and $x_2$,
\begin{eqnarray}
| \text{A} \rangle &=& \frac{1}{\sqrt{2}}( x_2 |j^z=2 \rangle + \overline{x_2} | j^z=-2 \rangle) + x_0 | j^z = 0\rangle
\nonumber \\
&+&\frac{1}{\sqrt{2}} ( x_1 |j^z=1\rangle + \overline{x_1} | j^z=-1 \rangle) 
\\
| \text{B} \rangle &=& \frac{1}{\sqrt{2}}( x_2 |j^z=2 \rangle +\overline{x_2} | j^z=-2 \rangle) + x_0 | j^z = 0\rangle
\nonumber \\
&+&\frac{1}{\sqrt{2}} ( - x_1 |j^z=1\rangle - \overline{x_1} | j^z=-1 \rangle) 
\\
| \text{C}\rangle &=& \frac{1}{\sqrt{2}}( -x_2 |j^z=2 \rangle - \overline{x_2} | j^z=-2 \rangle) + x_0 | j^z = 0\rangle
\nonumber \\
&+&\frac{1}{\sqrt{2}} (- i x_1 |j^z=1\rangle + i \overline{x_1} | j^z=-1 \rangle)
\\
| \text{D}\rangle &=& \frac{1}{\sqrt{2}}( -x_2 |j^z=2 \rangle - \overline{x_2} | j^z=-2 \rangle) + x_0 | j^z = 0\rangle
\nonumber \\
&+&\frac{1}{\sqrt{2}} ( i x_1 |j^z=1\rangle - i \overline{x_1} | j^z=-1 \rangle) 
\end{eqnarray}
with $x_0 = \sqrt{1-|x_1|^2-|x_2|^2}$.
The order parameters are:
\begin{eqnarray}
\langle {\bf  j}_{\text{A}/\text{B} } \rangle & = & \pm m (u_1,u_2,0) ,
\\
\langle {\bf  j}_{\text{C}/\text{D}} \rangle & = & \pm m (-u_2,u_1,0) ,
\\
\langle Q_{\text{A}/\text{B}/\text{C}/\text{D}}^{3z^2} \rangle & = & q,
\\
\langle Q_{\text{A}/\text{B}}^{x^2-y^2} \rangle & = & \langle Q_{\text{C}/\text{D}}^{x^2-y^2} \rangle = \pm q'.
\end{eqnarray}

It is easy to see from the above order parameters that the orbital
configuration still has a two-sublattice structure as demanded by the
orbital-orbital interaction while the magnetic dipolar order has a
four-sublattice antiferromagnetic structure. One can think of this state
as breaking time reversal symmetry on top of a two-sublattice orbitally
ordered state by developing antiferromagnetic orders within each
sublattice.  States on sublattice A and B form a time-reversal pair and
states on sublattice C and D form another time-reversal pair. But it is
not a conventional antiferromagnetic state.  In fact, in this phase, the
magnetic dipolar moment of A or B sublattice is nearly perpendicular to
the magnetic dipolar moment of C or D sublattice.  Consistent with
strong magnetic anisotropy, the staggered magnetizations $\langle {\bf
  j}_{\text{A}}-{\bf j}_{\text{B}}+{\bf j}_{\text{C}} - {\bf
  j}_{\text{D}} \rangle$ and $\langle {\bf j}_{\text{A}}-{\bf
  j}_{\text{B}}-{\bf j}_{\text{C}} + {\bf j}_{\text{D}} \rangle$ are
oriented in a direction very close to $[110]$ lattice direction.

\subsubsection{Four-sublattice antiferromagnetic (``$\overline{\Delta}$'') state}
\label{sec:sec326}

In the intermediate $J'/J$ and the small $V/J$ regime, we find another
four-sublattice antiferromagnetic phase, the ``$\overline{\Delta}$''
state. In this regime, the electric quadrupolar interaction may be
neglected, and we can understand the state as arising due to the
competition between FM and AFM exchange interactions.  In the AFM100
phase, every site has eight NN AFM neighbors and four NN FM
neighbors. In the FM111 phase, every site has zero NN AFM neighbors and
twelve NN FM neighbors. In the ``$\overline{\Delta}$'', we find that,
for every site there is four NN AFM neighbors which is an intermediate
case compared to FM111 and AFM100 phase.  The ground state wavefunction
of this ``$\overline{\Delta}$'' phase is parametrized by two complex
numbers, $x_1$ and $x_2$,
\begin{eqnarray}
|\text{A} \rangle &=&  \frac{1}{\sqrt{2}}  ( x_2 | j^z =2\rangle  + \overline{x_2} | j^z =- 2\rangle) +x_0 | j^z =0 \rangle
\nonumber \\
&&
+\frac{1}{\sqrt{2} } ( x_1 | j^z =1\rangle  + \overline{x_1} | j^z =- 1\rangle  ) 
\\
|\text{B} \rangle &=& \frac{1}{\sqrt{2}}  ( x_2 | j^z =2\rangle  + \overline{x_2} | j^z =- 2\rangle)+x_0| j^z =0 \rangle
\nonumber \\
&&
 + \frac{1}{\sqrt{2}} ( -x_1 | j^z =1\rangle  -\overline{x_1} | j^z =- 1\rangle  )
\\
|\text{C} \rangle &=&  \frac{1}{\sqrt{2} } ( \overline{x_2} | j^z =2\rangle  + x_2 | j^z =- 2\rangle) +x_0| j^z =0 \rangle 
\nonumber \\
&&
 + \frac{1}{\sqrt{2}} ( \overline{x_1} | j^z =1\rangle  + x_1 | j^z =- 1\rangle  ) 
\\
|\text{D} \rangle &=&  \frac{1}{\sqrt{2} }  ( \overline{x_2} | j^z =2\rangle  + x_2 | j^z =- 2\rangle)+x_0| j^z =0 \rangle
\nonumber \\
&&
+ \frac{1}{\sqrt{2} } ( -\overline{x_1} | j^z =1\rangle  -x_1 | j^z =- 1\rangle  )
\end{eqnarray}
with $x_0 =\sqrt{1-|x_1|^2-|x_2|^2}$. As one can see from the
wavefunction, states on A and B sublattices form a time reversal pair
and states on C and D sublattices form another time reversal pair.  The
magnetic dipolar and quadrupolar orders of each sublattice are found to
have the following relation,
\begin{eqnarray}
\langle {\bf  j}_{\text{A}/\text{B} } \rangle & = & \pm m (u_1,u_2,0) ,
\\
\langle {\bf  j}_{\text{C}/\text{D}} \rangle & = & \pm m (u_1,- u_2,0) ,
\\
\langle Q_{\text{A}/\text{B}/\text{C}/\text{D}}^{3z^2} \rangle & = & q,
\\
\langle Q_{\text{A}/\text{B}/\text{C}/\text{D}}^{x^2-y^2} \rangle & = &  q'.
\end{eqnarray}

Although both ``$\Delta$'' and ``$\overline{\Delta}$'' states are
four-sublattice states, they are actually different phases.  For
instance, unlike in the phase ``$\Delta$'' discussed in last section,
the quadrupole moments in the ``$\overline{\Delta}$'' phase are uniform.
Intuitively, this is because the orbital-orbital interaction is not
large enough to induce a two-sublattice structure in this regime.

\subsubsection{Two-sublattice quadrupole (spin nematic) state}
\label{sec:sec327}

With an intermediate $J'/J$ and a large $V/J$, the orbital-orbital
interactions dominate over all other interactions in the
Hamiltonian. Therefore the ground state reduces to quadrupolar phase
found in Sec.~\ref{sec:sec31}. The full set of order parameters is
\begin{eqnarray}
\langle {\bf j}_{\text{A} / \text{B} }\rangle & = & 0 \nonumber
\\
\langle  Q_{\text{A} / \text{B} }^{3z^2}  \rangle & = & 0 \nonumber 
\\
\langle Q_{\text{A} / \text{B}} ^{x^2- y^2 } \rangle & = & \pm q'\label{eq:8}
\end{eqnarray}
with $q'=2 \sqrt{3}$.

\subsubsection{$T=0$ Transitions}
\label{sec:transitions}

At the mean field level, the nature of the transitions between different
phases can be understood from the wavefunctions. If the wavefunction of
one phase can be continuously tuned to that of the neighboring phase,
then the phase transition between these two phases may be
continuous. Otherwise, it is first order.  We find the possible
continuous transitions are FM111-``$\ast$'', ``$\ast$''-FM110,
FM110-Quadrupolar, Quadrupolar-``$\Delta$'' and
``$\overline{\Delta}$''-AFM100.  The remaining transitions are first
order at mean field level.  We do not discuss fluctuation effects here,
which would be required for a full understanding of the transitions.

\section{$T>0$ phases}
\label{sec:sec4}

In this section, we study the effects of thermal fluctuations on the
phase diagram for $T>0$.   We do this using standard Weiss mean field
theory, taking into account the symmetry structure of
the phases identified in Sec.~\ref{sec:sec3}.  The mean field method
produces self-consistent equations for the order parameters, which can
be solved, choosing the solution with minimal free energy, to obtain
the temperature dependence of physical quantities.  These equations
are generally sufficiently complicated that, even taking into account
symmetry conditions, only a numerical solution
is possible.  

Before presenting the numerical solution, we discuss a few aspects of
the phase diagram which can be understood analytically.  Specifically,
we consider the instabilities of the paramagnetic state on decreasing
the temperature from large to small values.  Several instabilities are
possible, which are signaled in the mean field calculations by the
appearance of a solution of a given symmetry $S$ at a temperature
$T_S$.  Mathematically, the temperature $T_S$ is defined by the
vanishing of the coefficient of the quadratic term in the order
parameter associated with the symmetry $S$ in the Landau free energy,
or equivalently, the divergence of the mean field susceptibility of
the $S$ order parameter.  The order parameters in question are the
magnetization vector and quadrupolar tensor at ${\bf p}={\boldmath 0}$
and at ${\bf p}={\boldmath Q}=2\pi(001)$ (note the four
sublattice states should be described by several ${\bf p}=2\pi(001)$
wavevectors).  In this way we obtain four temperatures
$T_m({\boldmath 0})$, $T_m({\boldmath Q})$, $T_Q({\boldmath 0})$,
$T_Q({\boldmath Q})$.   For a given value of the exchange parameters
($J,J',V$), the {\sl largest} of these temperatures will determine the
actual instability of the paramagnetic state, and thus the first type
of order which is encountered upon lowering the temperature.  In
principle, a first order transition could also occur at a temperature
higher than this, but the full mean field solution shows that this
does not occur except in a region where $J'/J, V/J \ll 1$.  

The instability temperatures may determined analytically.  One finds
\begin{eqnarray}
T_m ({\boldmath 0}) &=& \frac{6J'-2 J }{5} + \frac{\sqrt{2 J^2 - 20 J J' + 52 J'^2}} {5} \\
T_Q ({\boldmath 0}) &=& \frac{7}{60} (J-6J'+7V), \\
T_m ({\boldmath Q})   &=& \frac{2}{15} \left[J - 2J' 
    +r \cos \frac{\alpha}{3}\right], \\
T_Q  ({\boldmath Q})  &=&  \frac{7}{180} (-3J +18 J'+43 V). 
\end{eqnarray}
Here we defined
\begin{eqnarray}
  \label{eq:9}
  r & =& \sqrt{6J^2 - 20JJ'+32 J'^2}, \\
 \alpha&=& Arg \Big[5-72y+204y^2-160y^3 \nonumber \\
&+& i \sqrt{(r/J)^6 - (5-72y+204y^2-160y^3)^2} \Big]
%\tan^2 \alpha = \\
%&&  (191- 1440 y +3432 y^2 - 64 
%  y^3 - 7824 y^4 \nonumber \\
 % & & + 3480 y^5+7168y^6)/(5-72y+204y^2-160y^3)^2,\nonumber
\end{eqnarray}
and $y=J'/J$.
\begin{figure}[htp]
\includegraphics[width=8.0cm]{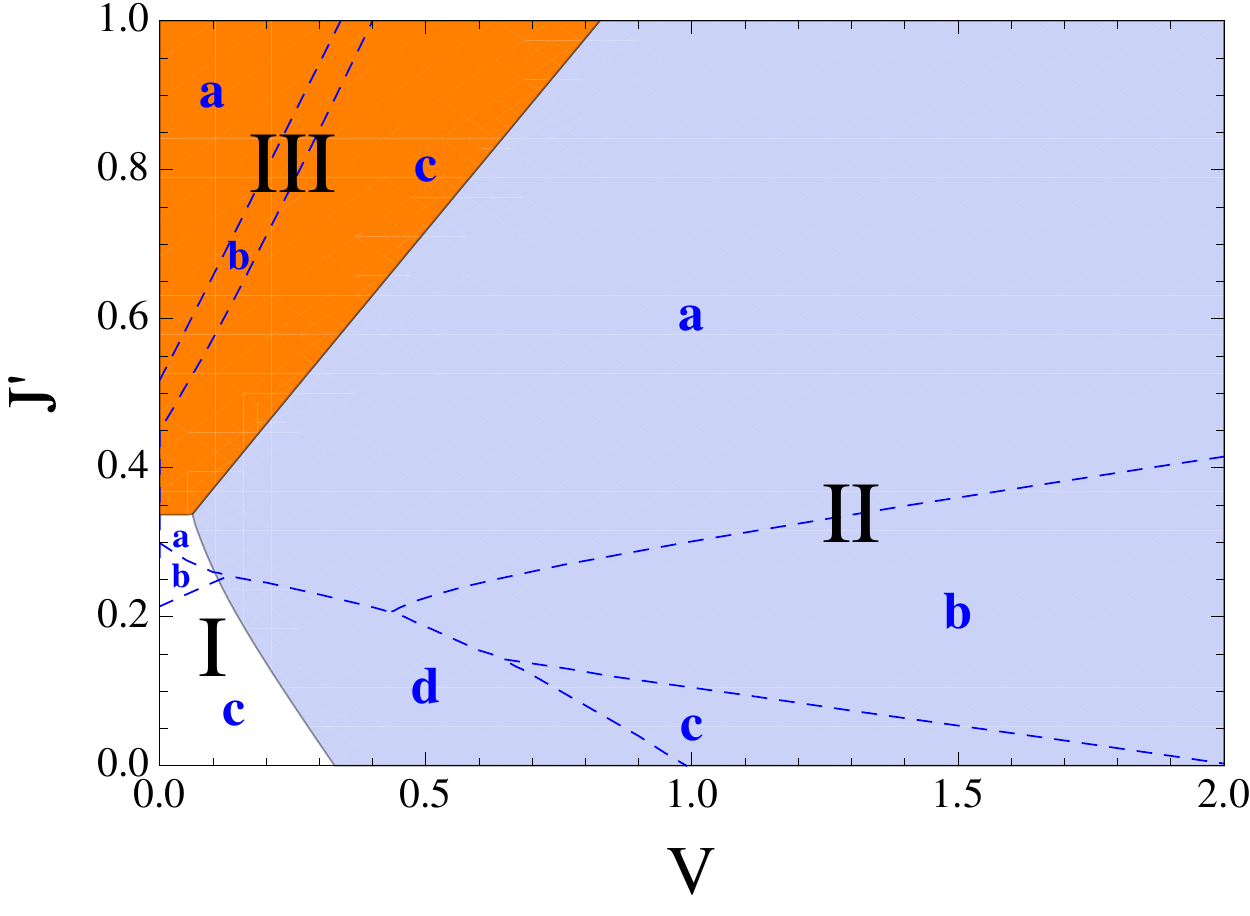}
\caption{ (Color online) $T>0$ phase diagram.  In Region I, (white)
  the system transitions directly from the high temperature
  paramagnetic phase to a magnetic state with ordering wavevector
  ${\bm Q}= 2\pi(001)$ at $T=T_m({\bm Q})$.  In region II (blue), the
  transition from the high temperature paramagnet is instead to a
  quadrupolar phase with ordering wavevector ${\bm Q}$, at $T=T_Q
  ({\bm Q})$.  In region III (red), the transition from the
  paramagnetic is to a ${\bf q}=0$ ferromagnetic state at $T=T_m ({\bm
    0})$.  The dashed curves are the boundaries of the ground
  state phases taken from Fig.~\ref{fig:pd1}. ``a, b, c, d'' label the
  low temperature phases of each region. $J=1$ in the phase diagram.}
\label{fig:pd2}
\end{figure}

In the parameter space depicted in Fig.~\ref{fig:pd1}, $T_Q ({\bf p} =
{\bf 0})$ is always smaller than the other three temperatures.  Thus
there is never an instability of the paramagnet to a uniform
quadrupolar ordered state.  Comparing the remaining three
temperatures, we find three distinct regions shown in
Fig.~\ref{fig:pd2}.  In region I, the highest transition temperature
is $T_m ({\boldmath Q})$, and magnetic order with an enlarged unit
cell sets in directly from the paramagnetic state.  In region II, the
highest transition temperature is $T_Q ({\boldmath Q})$ and
two-sublattice quadrupolar (spin nematic) occurs neighboring the
paramagnetic state.  In region III, the highest transition temperature
is $T_m ({\boldmath Q})$, and the paramagnetic phase undergoes a transition
directly to a ferromagnetic one.

On further lowering of temperature, additional phases may occur.  Full
mean-field calculations show that there are in fact are ten different
patterns of thermal evolution, indicated by the the different
sub-regions in Fig.~\ref{fig:pd2}.  We discuss this in further detail below.

\subsection{Region I}

\begin{figure}[htp]
\centering
\subfigure[  $\;J'=0.32J, V=0.05J$ in I$_a$]{\includegraphics[width=7cm]{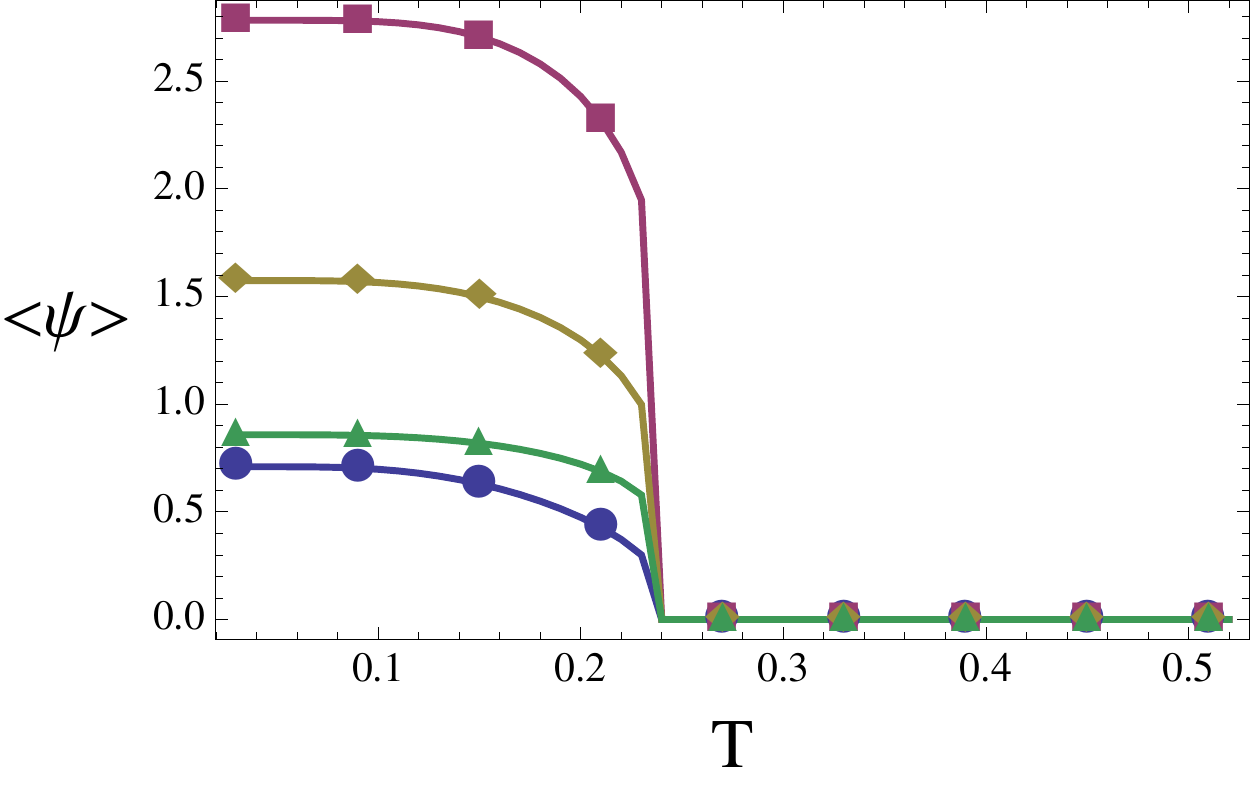}}
\label{fig:Ia}
\subfigure[ $\; J'=0.25J, V=0$ in I$_b$]{\includegraphics[width=7cm]{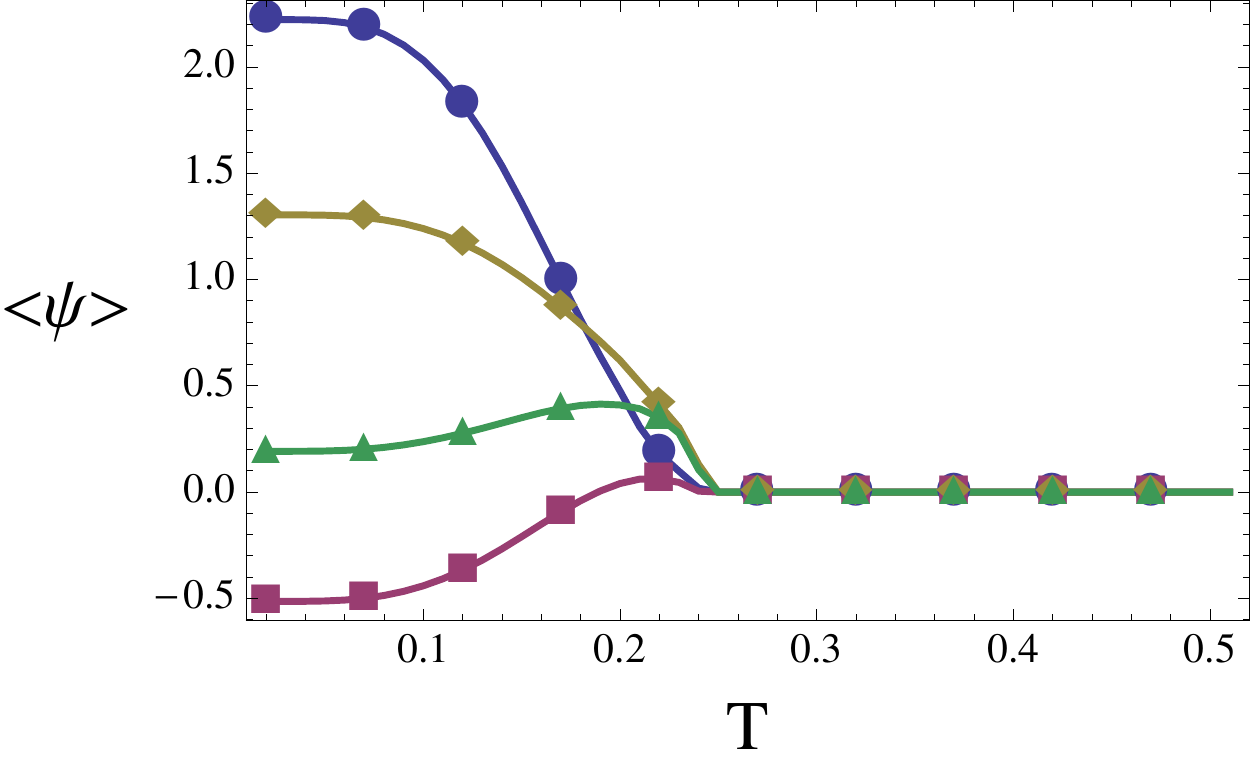}}
\label{fig:Ib}
\subfigure[ $\; J'=V=0.1J$ in I$_c$.]{\includegraphics[width=7cm]{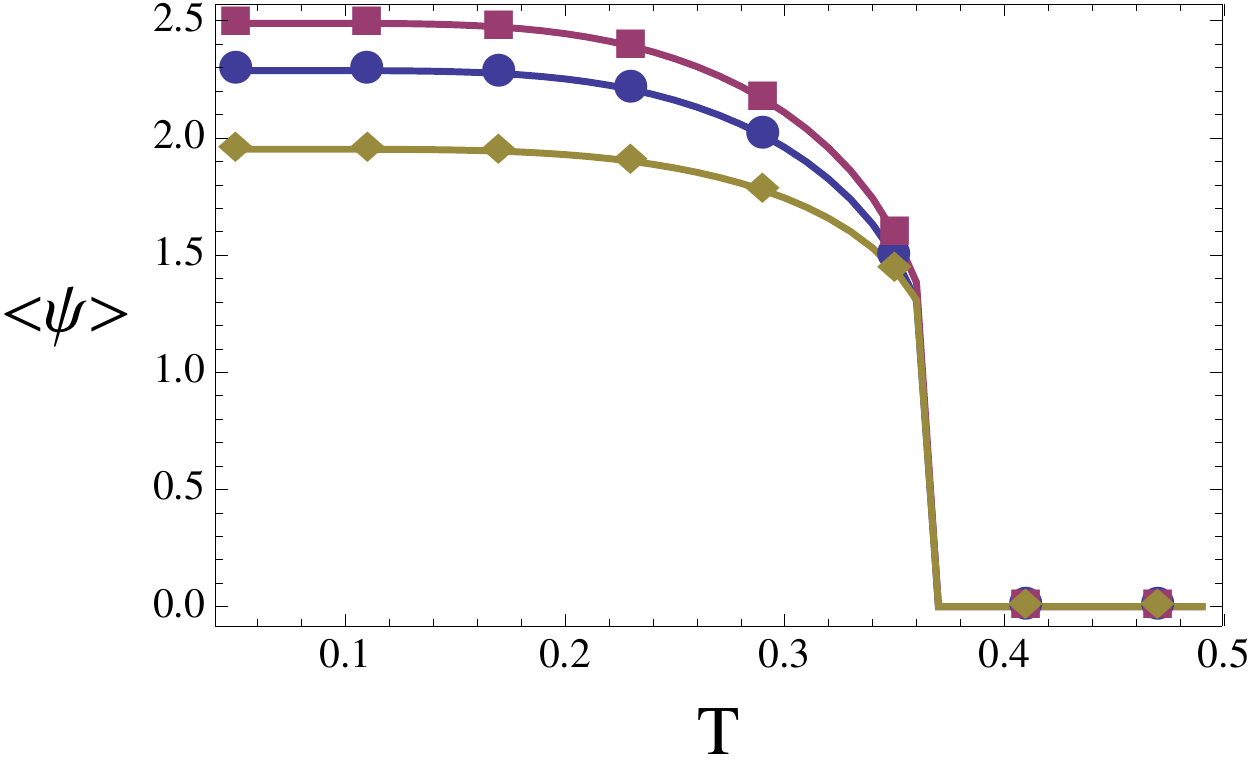}}
\label{fig:Ic}
\caption{(Color online)  Order parameters plotted for three subregions of region I:
(a) 
Sqaure (red) $(\langle Q_{\text A}^{x^2-y^2}-Q_{\text B}^{x^2-y^2}\rangle)/2$,
ball (blue) $\langle Q_{\text{A}}^{3z^2}  + Q_{\text{B}}^{3z^2} \rangle/2 $,
diamond (yellow)  $|\langle {\bf j}_{\text A}+ {\bf j}_{\text B}\rangle|/2$, and
triangle (green) $|\langle {\bf j}_{\text A} -  {\bf j}_{\text B}\rangle|/2$.
(b) 
Sqaure (red) $ \sum_{i=\text{A,B,C,D}}\langle Q_i^{3z^2} \rangle $, 
ball (blue) $ \sum_{i=\text{A,B,C,D}}  \langle Q_i^{x^2-y^2} \rangle $, diamond (yellow)  
$|\langle {j}^x_{\text{A} } -{ j}^x_{\text{B}} + j^x_{\text{C}} - j^x_{\text{D}}  \rangle|/4$, and
triangle (blue) $|\langle {j}^y_{\text{A} } -{ j}^y_{\text{B}} - j^y_{\text{C}}  + j^y_{\text{D}}  \rangle|/4$.
(c) 
Sqaure (red) $ \langle Q^{x^2-y^2}_{\text{A}} + Q^{x^2-y^2}_{\text{B}}\rangle /2$, 
ball (blue) 
$\langle Q^{3z^2}_{\text{A}} + Q^{3z^2}_{\text{B}} \rangle/2$, and diamond (yellow)  
$|\langle {\bf j}_{\text{A} } -{\bf j}_{\text{B}}  \rangle|/2$.
}
\label{fig:regionI}
\end{figure}

In region I, the system has a direct transition from the high
temperature paramagnetic phase to low temperature magnetically ordered
phases specified by the dashed curves in Fig.~\ref{fig:pd2}.  At mean
field level, the transitions to FM110 phase in region I$_a$ and AFM100
phase in region I$_c$ are found to be first order, while a continuous
transition to the four-sublattice ``$\overline{\Delta}$'' phase is
observed in region I$_b$ (see Fig.~\ref{fig:regionI}).   

\subsection{Region II}
\label{sec:sec42}

\begin{figure}[htp]
\centering
\subfigure[ $\; J'=0.6 J, V=0.8J$ in II$_a$]{\includegraphics[width=7cm]{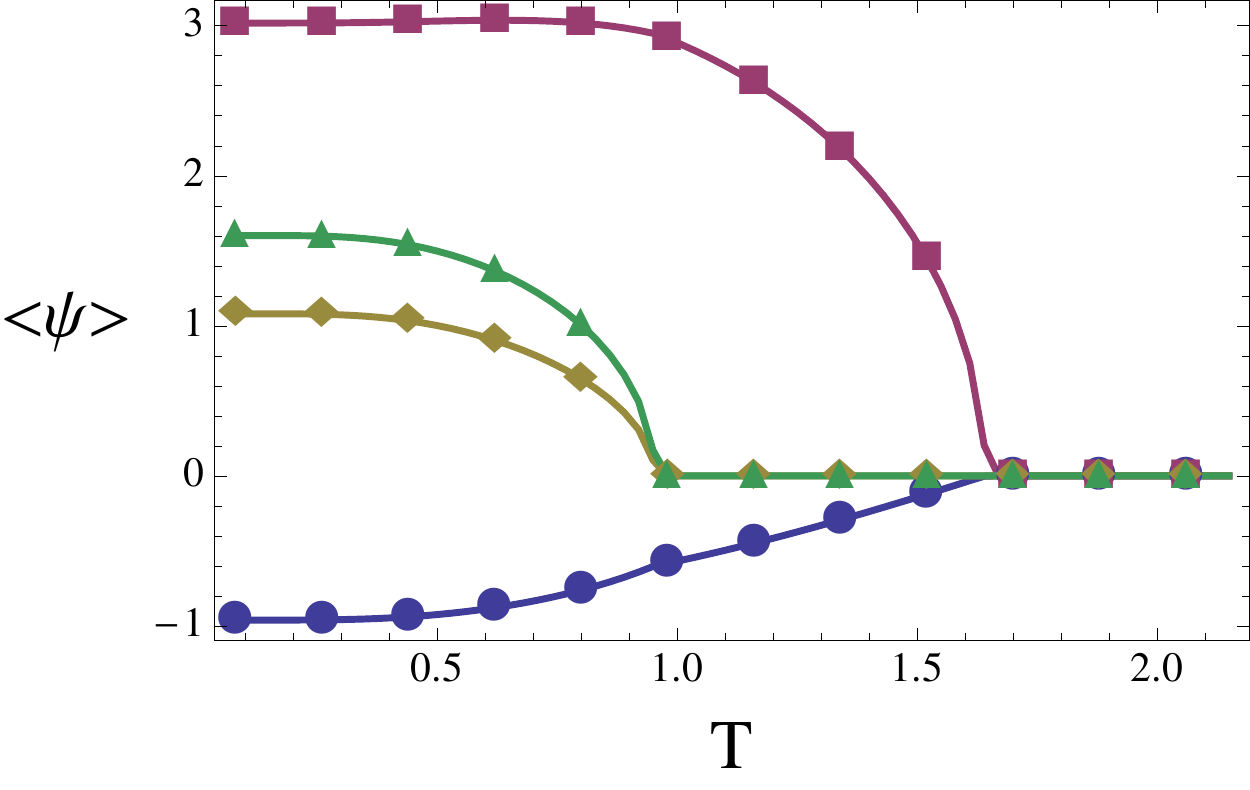}}
\label{fig:IIa}
\subfigure[$\; J'=0.2 J, V=0.9J$ in II$_b$. ]{\includegraphics[width=7cm]{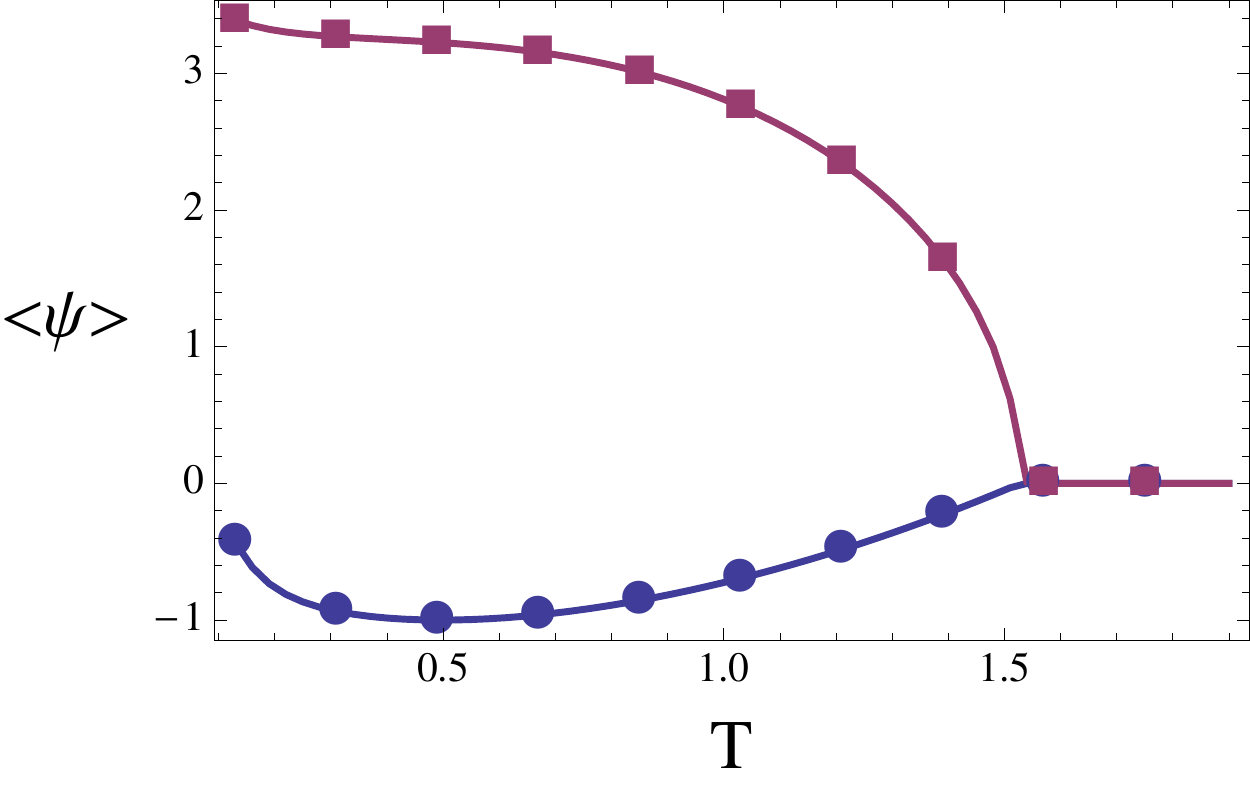}}
\label{fig:IIb}
\subfigure[ $\;J'=0.05J, V=1.0J$ in II$_c$. ]{\includegraphics[width=7cm]{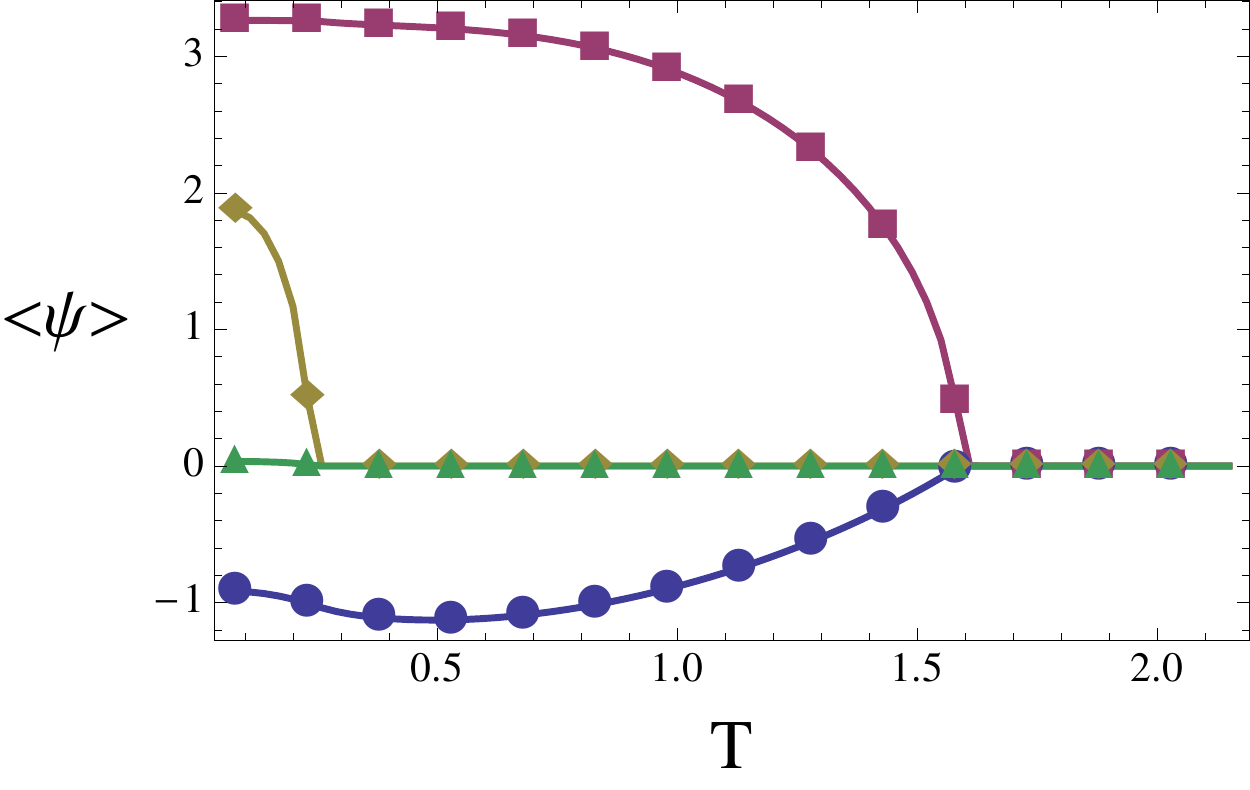}}
\label{fig:IIc}
\caption{(Color online) Order parameters plotted in four subregions of region II:
(a)
square (red) $\langle Q^{x^2-y^2}_{\text A} - Q^{x^2-y^2}_{\text B} \rangle /2$,
ball (blue) $\frac{1}{2}\langle Q_{\text A}^{3z^2} + Q_{\text B}^{3z^2} \rangle$,
triangle (green) $ |\langle {\bf j}_{\text A} + {\bf j}_{\text B}  \rangle |/2$, and
diamond (yellow)  $ |\langle {\bf j}_{\text A} - {\bf j}_{\text B}  \rangle |/2 $.
(b) square (red) $\langle Q^{x^2-y^2}_{\text A} - Q^{x^2-y^2}_{\text B} \rangle /2$ and
ball (blue) $\frac{1}{2}\langle Q_{\text A}^{3z^2} + Q_{\text B}^{3z^2} \rangle$
(c)  square (red) $\langle Q^{x^2-y^2}_{\text A} + Q^{x^2-y^2}_{\text B} - Q^{x^2-y^2}_{\text C} 
- Q^{x^2-y^2}_{\text D} \rangle /4$,
ball (blue) $\sum_{i=\text{A,B,C,D} } \langle Q^{3z^2}_{i}  \rangle /4$, and
diamond (yellow)  $\langle Q^{x^2-y^2}_{\text A} + Q^{x^2-y^2}_{\text B} \rangle /2$,
upper triangle (green)  $|\langle {\bf j}_{\text A} - {\bf j}_{\text B}\rangle |/2$, and
down triangle (blue)   
$|\langle {\bf j}_{\text A} + {\bf j}_{\text B} \rangle|/2$.
}
\label{fig:regionIIabc}
\end{figure}

\begin{figure}[htp]
\includegraphics[width=7.0cm]{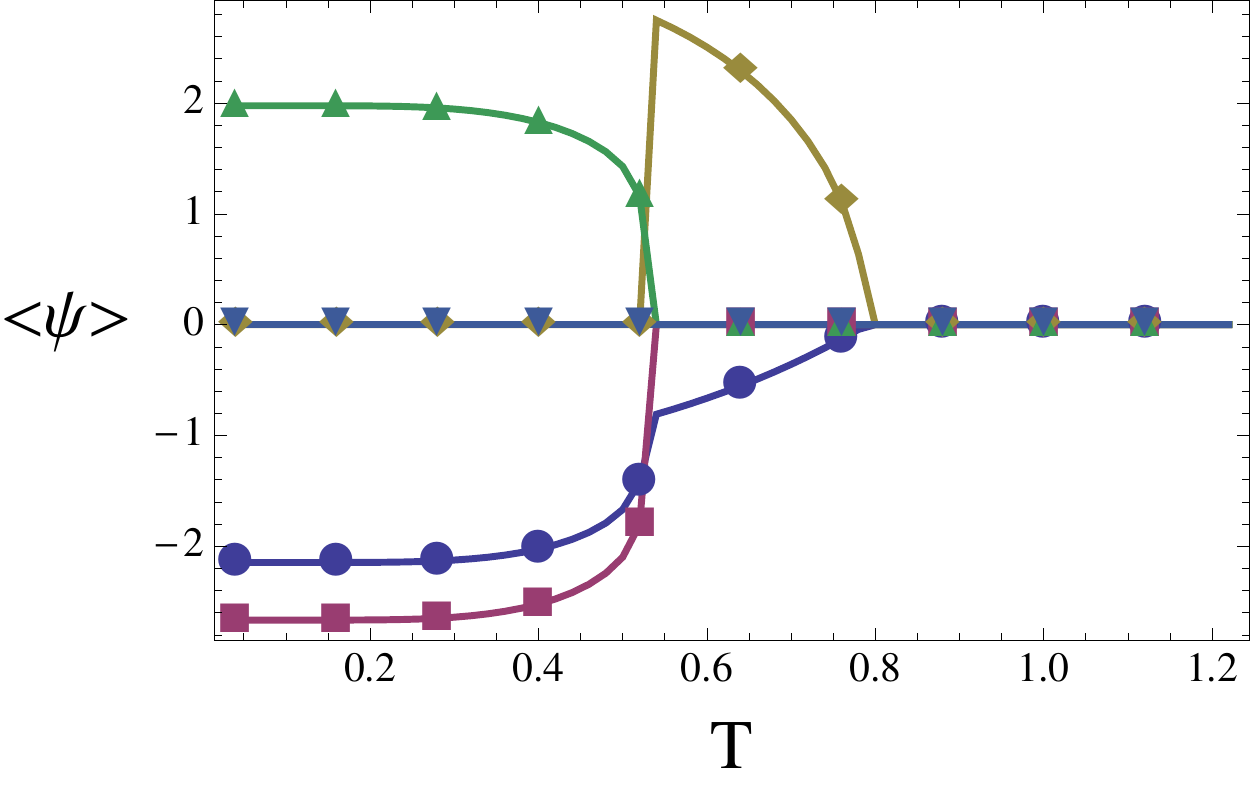}
\caption{(Color online) Order parameters in region II$_d$ with $ J'=0.1 J, V=0.5J$:
square (red) $\langle Q^{x^2-y^2}_{\text A} - Q^{x^2-y^2}_{\text B} \rangle /2$,
ball (blue) $\frac{1}{2}\langle Q_{\text A}^{3z^2} + Q_{\text B}^{3z^2} \rangle$,
diamond (yellow) $\langle j^x_{\text A} - j^x_{\text B} + j^y_{\text C} - j^y_{\text D} \rangle /4$, and
triangle (green) $\langle j^y_{\text A} - j^y_{\text B} - j^x_{\text C} + j^x_{\text D} \rangle /4$.
}
\label{fig:IId}
\end{figure}

In region II, there is a broad ${\bf p} = 2\pi (001)$ quadrupolar
phase in the intermediate temperature.  Unlike the spin nematic state
in the ground state phase diagram, this intermediate temperature
quadrupolar phase is actually a biaxial spin nematic state in which
the quadrupole moment $\langle Q^{\mu \nu}_i\rangle$ has three
distinct eigenvalues.   At mean field level, the transition from the PM
phase to quadrupolar phase is continuous (see
Fig.~\ref{fig:regionIIabc} and Fig.~\ref{fig:IId}).  Beyond mean field
theory, this transition is believed to be in a three dimension $O(3)$
universality class.\cite{PhysRevB.82.174440}

As mentioned previously, the low temperature phases (FM110 in region
II$_a$, phase ``$\Delta$" in region II$_c$) can be regarded as further
breaking the time reversal symmetry coming from the quadrupolar phase
at intermediate temperature. This transition is found to be continuous
at mean field level. The symmetry breaking associated with this
transition can be described by several Ising order parameters (uniform
and staggered magnetization).  This transition may be
continuous beyond mean field.\cite{PhysRevB.82.174440}

The transition from quadrupolar phase to AFM100 in region II$_d$ is
found to be strongly first order in mean field theory
(see Fig.~\ref{fig:IId}). This is easy to understand from a simple
symmetry analysis. The AFM100 phase has a uniform orbital
configuration and its symmetry is not a subgroup of the quadrupolar
phase.  Hence, it is almost impossible for the transition to be
continuous.

\subsection{Region III}
\label{sec:sec43}

In region III, the intermediate temperature phase is FM111. The
transition from PM to FM111 is found to be continuous at mean field
leavel.  The transition from FM111 to low temperature phases (FM110
and phase ``$\ast$'') is also found to be continuous in our mean field
analysis.  The transition from FM111 to phase FM110 can be simply
described by an Ising valuable $m^z$.  Similarly, the transition from
FM111 to phase ``$\ast$'' can also be described by an Ising valuable
which is the staggered magnetization.  Thus, these two transitions
could be continuous beyond mean field.

\begin{figure}[htp]
\centering
\subfigure[$\; J'=0.8J, V=0.1J$ in III$_a$]{\includegraphics[width=7cm]{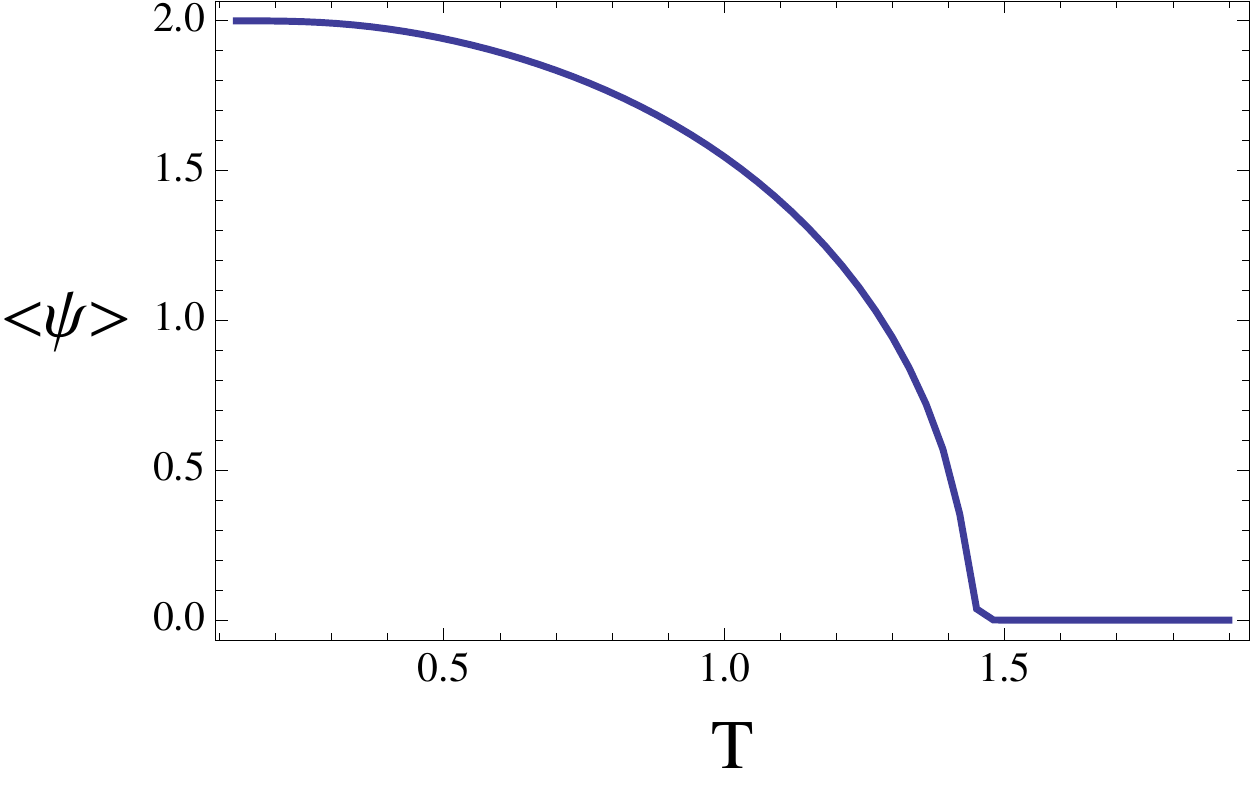}}
\label{fig:IIa}
\subfigure[ $\; J'=0.8J, V=0.22J$ in III$_b$]{\includegraphics[width=7cm]{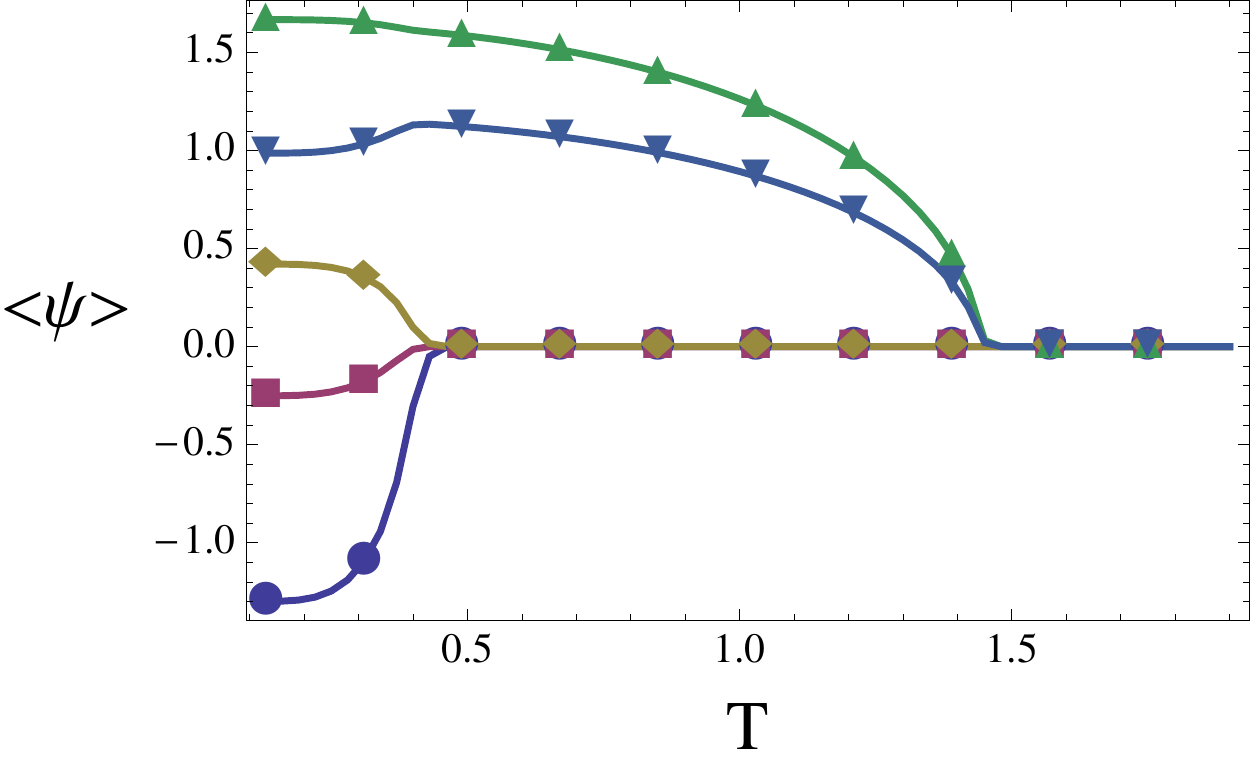}}
\label{fig:IIb}
\subfigure[$\; J=0.8J, V=0.5J$ in III$_c$]{\includegraphics[width=7cm]{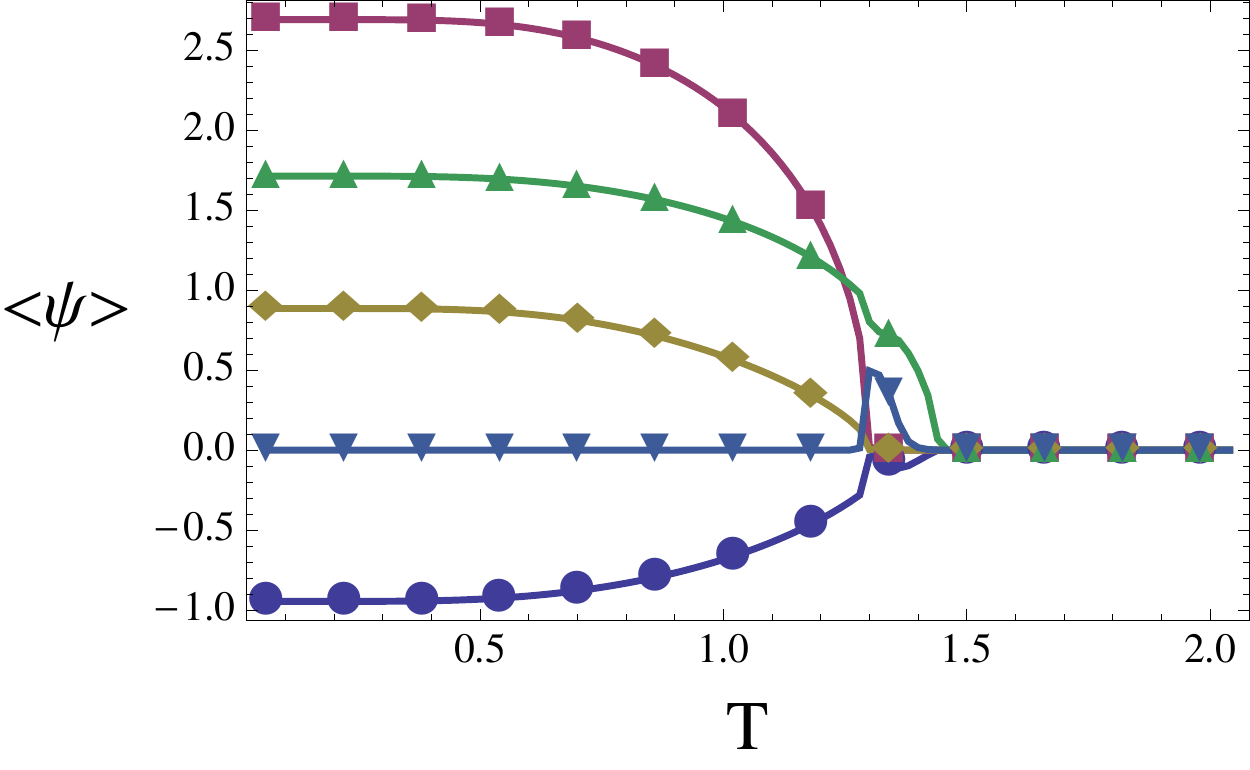}}
\label{fig:IIc}
\caption{(Color online) Order parameters plotted in three subregions of region III:
(a)  uniform magnetization $|\langle {\bf j}_i \rangle|$
(b)  square (red) $\langle Q^{3z^2}_{\text A} + Q^{3z^2}_{\text B} \rangle /2$,
ball (blue)  $\langle Q^{x^2-y^2}_{\text A} - Q^{x^2-y^2}_{\text B} \rangle /2$,
diamond (yellow)  $|\langle {\bf j}_{\text A} - {\bf j}_{\text B} \rangle/2  |$,
upper triangle (green)  $|\langle {\bf j}_{\text A}^{\perp}+ {\bf j}_{\text B}^{\perp} \rangle/2  |$, and
down triangle (blue)  $|\langle {\bf j}_{\text A}^{z}+ {\bf j}_{\text B}^{z} \rangle/2  |$.
(c)  
square (red)  $\langle Q^{x^2-y^2}_{\text A} - Q^{x^2-y^2}_{\text B} \rangle /2$,
ball (blue)  $\langle Q^{3z^2}_{\text A} + Q^{3z^2}_{\text B} \rangle /2$,
diamond (yellow)  $ |\langle {\bf j}_{\text A} - {\bf j}_{\text B} \rangle/2  |$,
upper triangle (green)   $|\langle {\bf j}_{\text A}^{\perp}+ {\bf j}_{\text B}^{\perp} \rangle/2  |$,
and down triangle (blue)   $| \langle {\bf j}_{\text A}^{z} + {\bf j}_{\text B}^{z}   \rangle /2  |$ .}
\label{fig:regionIII}
\end{figure}

\subsection{Magnetic susceptibility}
\label{sec:sec44}

In this subsection, we discuss the magnetic response at $T>0$. We
focus particularly on the intriguing non-magnetic quadrupolar phase of
region II in Fig.~\ref{fig:pd2}.  The magnetic response is found to be
an important indicator for the quadrupolar ordering transition.

\begin{figure}[htp]
\centering
\subfigure[$\; J'=0.4J, V=0.5J$ in II$_a$]{\includegraphics[width=7cm]{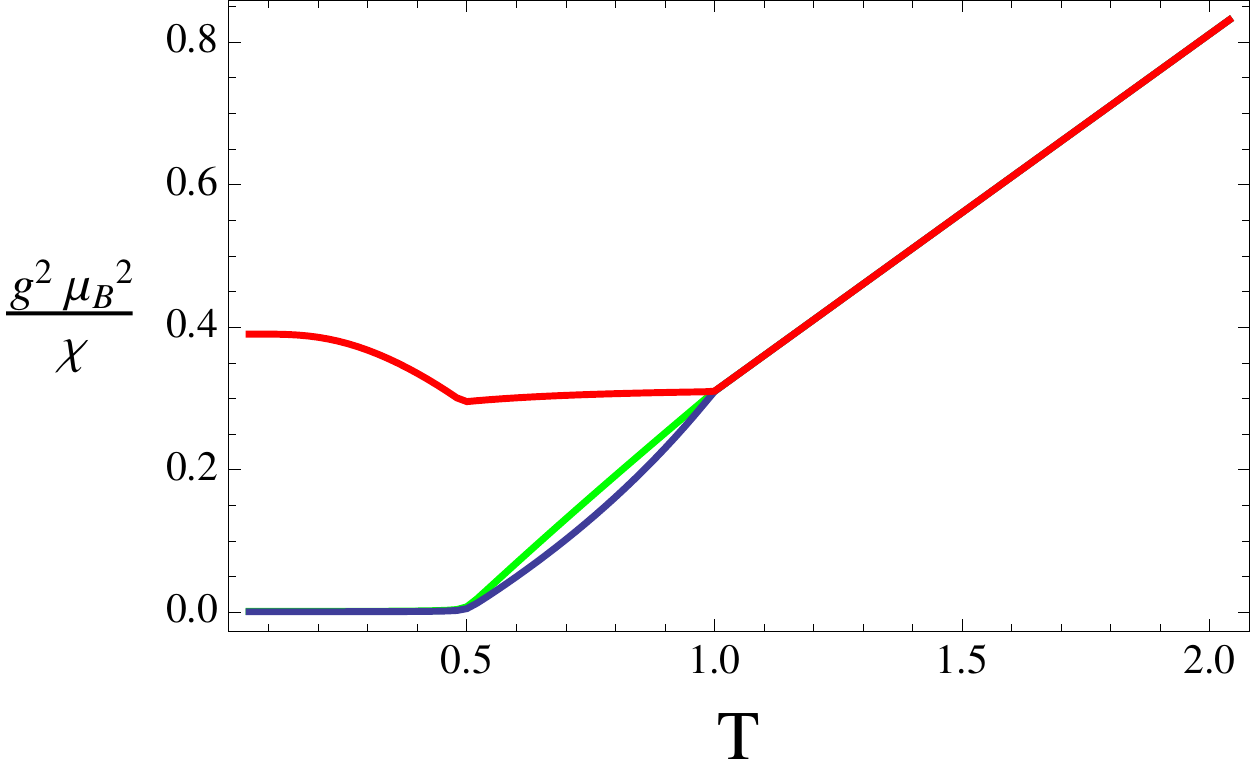}}
\label{fig:mag1}
\subfigure[$\; J'=0.2J, V=0.55J$ in II$_b$]{\includegraphics[width=7cm]{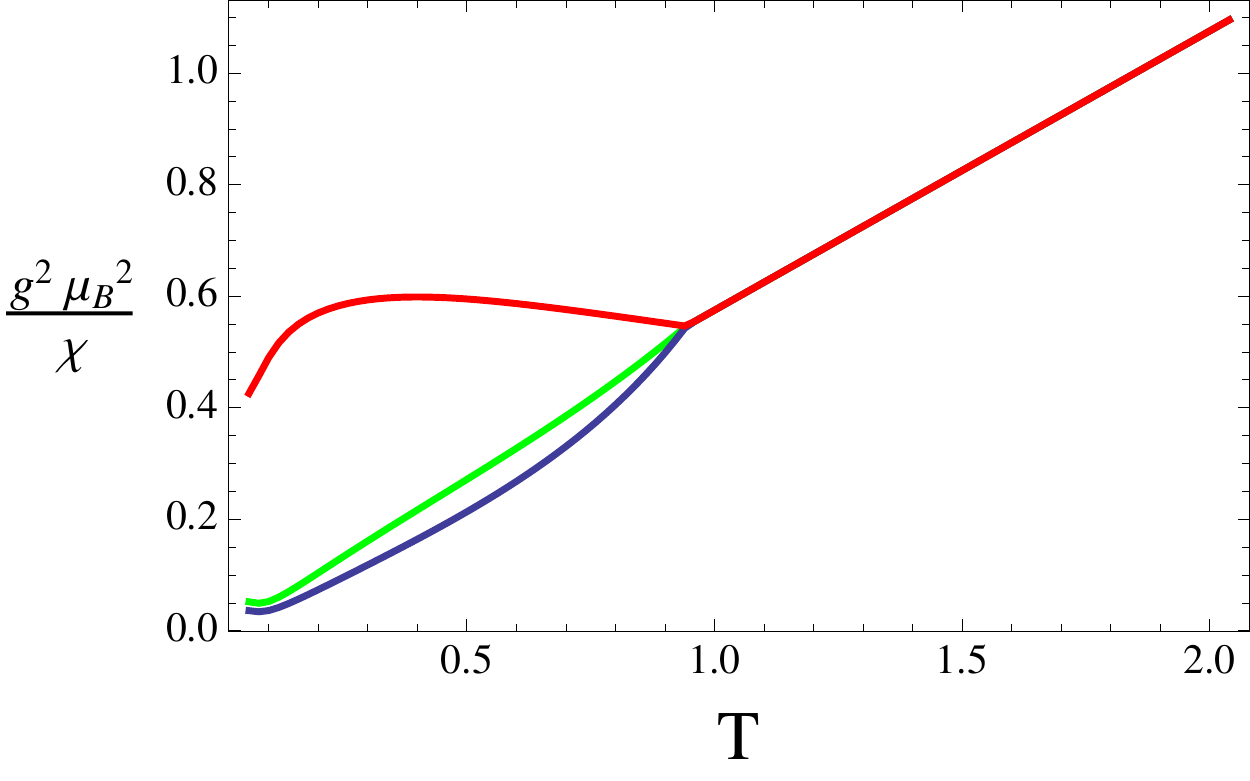}}
\label{fig:mag2}
\subfigure[$\; J'=0.05J, V=1.0J$ in II$_c$]{\includegraphics[width=7cm]{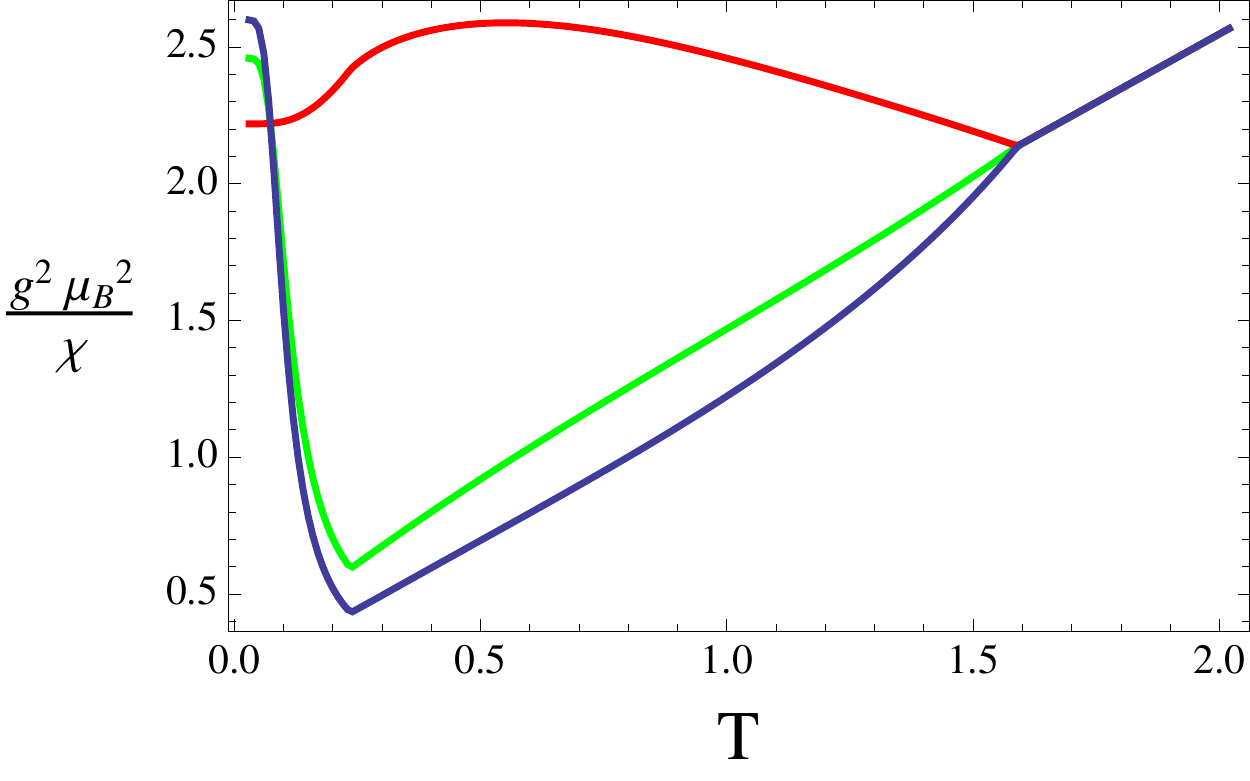}}
\label{fig:mag3}
\subfigure[$\; J'=0.1J, V=0.7J$ in II$_d$]{\includegraphics[width=7cm]{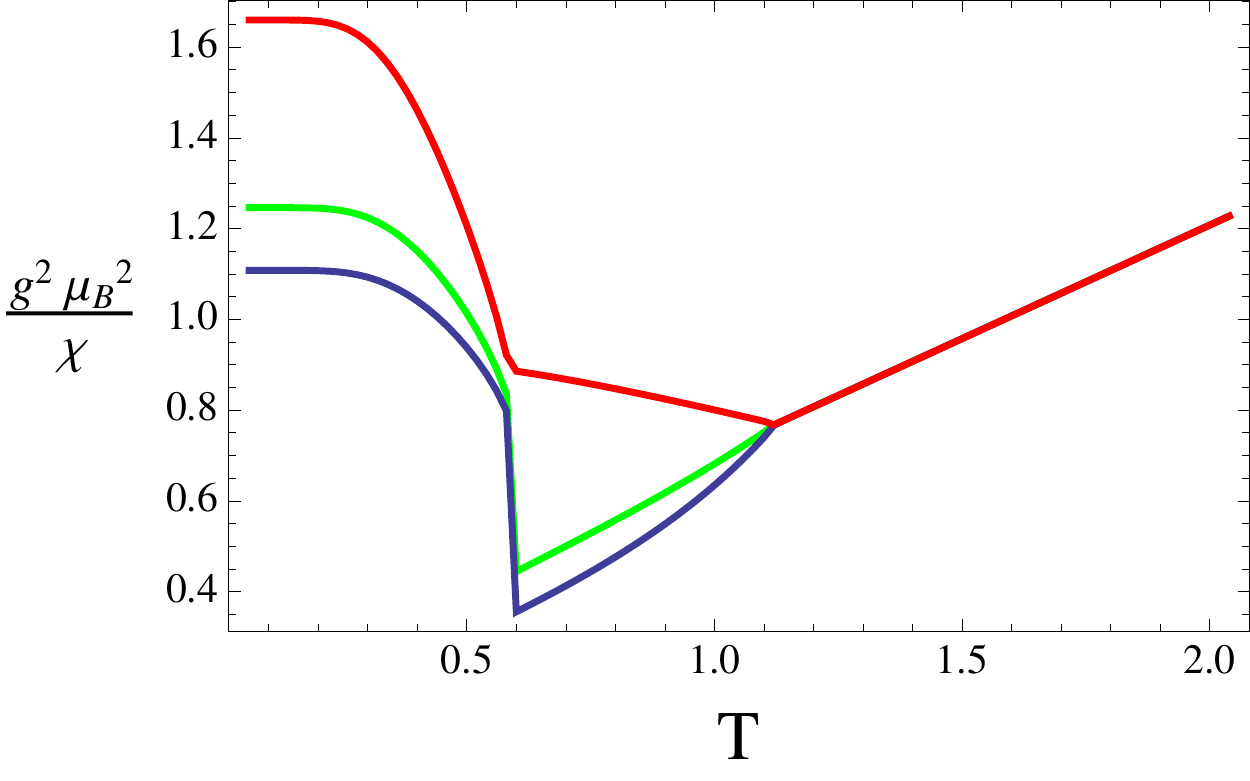}}
\label{fig:mag4}
\caption{(Color online) Inverse magnetic susceptibility of different subregions of region II in Fig.~\ref{fig:pd2}.
Blue (lower) curve: $1/\chi_{xx}$, red (upper) curve: $1/\chi_{zz}$, green (middle) curve: $1/\chi_{\text powder}$.}
\label{fig:magII}
\end{figure}

The general features, observed in Fig.~\ref{fig:magII}, are as follows.  
At high temperatures, the magnetic susceptibility $\chi$ obeys the
Curie-Weiss law, $\chi^{-1} \sim A(T-\Theta_{cw})$, when $T\gg
\Theta_{cw}$.  The Curie-Weiss temperature is readily obtained from a
high-temperature series expansion:
\begin{equation}
  \Theta_{cw} = \frac{-17 J + 66 J'}{25}.
\label{eq:cw}
\end{equation} 
From the mean-field solution, we obtain the susceptibility at lower
temperature.  At the quadrupolar ordering transition, a cusp in $\chi$
is observed (see Fig.~\ref{fig:magII}). This cusp separates the true
Curie-Weiss regime of the PM phase from a second Curie-Weiss regime at
intermediate temperatures.  The existence of two Curie-Weiss regimes
can be understood as due to the remaining magnetic degeneracy of the
quadrupolar phase.  Specifically, the intermediate quadrupolar phase
partially lifts the five-fold spin-orbital degeneracy, giving rise to
a local doublet which is a time-reversal pair. One should note that
this pair is not, however, a Kramer's pair.  This doublet is
responsible for the Curie-Weiss behavior at the intermediate
temperature regime.  Looking in more detail, since the spin nematic
phase at intermediate temperature has a tetragonal symmetry, we obtain
two different susceptibilities: parallel to the wavevector $p = 2\pi
(001)$ ($\chi_{zz}$) and normal to it ($\chi_{xx}=\chi_{yy}$).  As the
temperature is lowered further and magnetic order develops, this is
reflected in additional features in the susceptibility.

%*************************************************************************************************

\section{Discussion}
\label{sec:sec5}

% summary

In this paper, we introduced and analyzed a spin-orbital model to
describe the localized electrons in a 4d$^2$ or 5d$^2$ configuration
on an fcc lattice, in which strong spin-orbit coupling and the
three-fold degeneracy of the two-electron orbital states combine to
induce a local effective $j=2$ moment.  Nearest neighbor
antiferromagnetic and ferromagnetic exchange interactions and electric
quadrupolar interaction were included in the model Hamiltonian. We
obtained the ground state and finite temperature phase diagrams by 
Weiss mean field theory.   Seven different ground states (or low
temperature) phases were found.  In a large portion of the parameter
space, the system develops a two-sublattice structure of orbital
configuration which is driven by the diagonal orbital-orbital
interaction.  Most interestingly, a non-magnetic spin nematic ground
state occurs in the ground phase diagram and extends to a large
portion of the phase diagram at finite temperature.  Moreover, we find
ten different ways for the system to evolve from the high temperature
paramagnetic phase to the seven low temperature phases.

Our theory has provided numerous preditions for experiment.  For the
magnetic ordered phases which break time-reversal symmetry, neutron
scattering, NMR and/or magnetization measurements can probe the
magnetic structure.  For the spin nematic phase, similar to the one
electron case discussed in Ref.~\onlinecite{PhysRevB.82.174440}, the
magnetic quadrupole order is expected to induce a distortion of the
lattice and lower the crystal symmetry. Specifically, the quadrupolar
phase corresponds to the tetragonal space group $P4_2/mnm$ (number
136),\cite{PhysRevB.82.174440}, distinct from the the cubic space
group $Fm\bar{3}m$ of the high temperature phase. If this distortion
leads to a measurable effect,  high resolution x-ray scattering
should be able to identify the spin nematic order.  The low temperature
spin nematic state may also be identified directly from measurements
of the orbital state by resonant x-ray scattering or x-ray
reflectometry, which could be compared with the theoretical
wavefunctions in Sec.~\ref{sec:sec31}.

% material 

Now we discuss the specific materials which have been studied
experimentally to date.  We start from
Ba$_2$CaOsO$_6$.\cite{yamamura:jssc2006} It retains the cubic
$Fm\bar{3}m$ structure down to $17\text{K}$. The Curie-Weiss
temperature of this material is $-157\text{K}$.  The magnetic moment
$1.61 \mu_{\text{B}}$ is much smaller than the spin only contribution
$2.83 \mu_{\text{B} }$ for spin $S=1$, but close to our prediction
$1.25 \mu_{\text{B}}$ based on strong SOC in Sec.~\ref{sec:sec21}.
The deviation can be understood from the effect of hybridization of
the Os $d$ orbitals with the oxygen $p$ orbitals, which increases the
local magnetic moment.\cite{PhysRevB.82.174440} Both magnetic
susceptibility and specific heat measurement find a single
antiferromagnetic phase transition at $T_{\text N} = 51\text{K}$.
According to our theory, a single magnetic transition correponds to
region I in Fig.~\ref{fig:pd2}. Since the low temperature phase of
Ba$_2$CaOsO$_6$ is antiferromagnetic, then the AFM100 and
``$\overline{\Delta}$'' phases are consistent with magnetization
measurements.  The enlarged unit cell and detailed orientation of
magnetic moments predicted here for these phases should provide
targets for future neutron scattering measurements.

The structure of La$_2$LiReO$_6$ was observed to be monoclinic (space
group $P2_1/n$).\cite{PhysRevB.81.064436} This material has a
Curie-Weiss temperature $-204\text{K}$, indicating a large
antiferromagnetic exchange. The magnetic moment is
$1.97\mu_{\text{B}}$, the smallnest of which suggests the important of
strong SOC. Magnetic susceptibility, neutron diffraction and $\mu$SR
did not find magnetic long range order down to $2\text{K}$, pointing
to a possible quantum spin liquid phase in this material.    Since this
crystal structure of this material deviates strongly from a cubic one,
our predictions based on the cubic structure are not applicable.
Instead, some splitting of the $j=2$ manifold should be taken into
account. Because this is a non-Kramers ion,
one could imagine a trivial non-magnetic ground state at the single
ion level.  However, we note that in Ref.\onlinecite{PhysRevB.81.064436}
substantial differences were observed between zero field and field
cooled samples below 50K, which was argued to be evidence
against a single-ion singlet.  If the local ground state is instead a
doublet, an exotic ground state could be favored, as proposed for the isostructural
material La$_2$LiMoO$_6$.\cite{arxiv1011} More theoretical and
experimental study of the crystal field splitting and multiplet
structure is required to further understand the ground state.

Ba$_2$YReO$_6$ is another material with a cubic crystal structure
(space group $Fm\bar{3}m$).\cite{PhysRevB.81.064436} The Curie-Weiss
temperature is $-616\text{K}$, suggestinga a predominant
antiferromagnetic exchange. The magnetic moment $1.93 \mu_{\text{B}}$
is consistent with the picture of strong SOC. The magnetic
susceptibility data clearly suggests two transitions and shows two
Curie regimes.  The first transition is at $\sim 150 \text{K}$ and the
second (spin freezing) transition is at $\sim 50 \text{K}$. The second
Curie regime appears at the intermediate temperatures between
$50\text{K}$ and $150\text{K}$. Neutron diffraction shows the absence
of detectable magnetic Bragg peaks.  $\mu$SR relaxation data observes
spin freezing at low temperature.  We may speculate that the spin
freezing results from disruption by defects of an ordered phase that
would otherwise occur in an ideal sample.  From the existence of two
Curie regimes, we postulate that Ba$_2$YReO$_6$ corresponds to region
II in the phase diagram Fig.~\ref{fig:pd2}. This would identify the
intermediate temperature phase as a spin nematic.  Interestingly,
$\mu$SR measured below 100K found evidence for two spin components,
which may be consistent with the two-sublattice nature of the
quadrupolar/spin-nematic phase.  Let us consider the low temperature
phase.  Given the very large negative Curie-Weiss temperature, the
ferromagnetic exchange is likely weak in Ba$_2$YReO$_6$.  Hence,
comparing with Fig.~\ref{fig:pd1}, the natural low temperature phases
are AFM100, the four-sublattice ``$\Delta$'' phase or spin nematic
phase.  Characterization of the type of disorder in this material
would be helpful in further elucidating the physics.  It would also be
interesting to more directly attempt to detect the proposed
quadrupolar order experimentally in the intermediate temperature
phase.  

This paper (and the related study in Ref.\onlinecite{PhysRevB.82.174440})
provide a theoretical framework to understand the magnetism and
orbital physics of this class of materials.  Looking to the future,
there is considerable room for refinement of the theory.  It would be
useful and interesting to include the Jahn-Teller effect, and to
develop some microscopic understanding (perhaps from {\sl ab-initio}
calculations) of the crystal field splittings in non-cubic materials.
Within the present model, more studies of thermal and quantum
fluctuations beyond mean-field and spin-wave approaches would be
desirable.  An understanding of the types of disorder in these
materials and their effects on the magnetism is also needed.
Hopefully continued pursuit and refinement of measurements
on this interesting class of materials will motivate further
theoretical work on these and other points. 

%Vanadium spinel also belongs to the type of system studied here.

%One drawback of our theory is that we neglect the Jahn-Teller effect which is always present in the 
%orbitally degenerate systems. The Jahn-Teller effect is usually considered to be weak in $t_{2g}$ system.
%An extension of our work is to include the Jahn-Teller effect explictly and analyze its competition 
%with the orbital-orbital interactions from exchange and Coulomb origin. 

\acknowledgments

This work was supported by the DOE through Basic Energy Sciences grant
DE-FG02-08ER46524.  LB’s research facilities at the KITP were
supported by the National Science Foundation grant NSF PHY-0551164.
GC was supported by the National Science Foundation grant NSF-PFC.

\bibliography{ref}

\end{document}